\documentclass[twocolumn]{aastex631}
\usepackage{graphicx}
\usepackage{bm}
\usepackage{amsmath}
\makeatletter
\def\fixmathlinenumbering{}
\makeatother
\usepackage{booktabs} 
\usepackage{makecell}
\usepackage{threeparttable}
\usepackage{multirow}

\shorttitle{Disk-induced IMBH growth}
\shortauthors{Wang et al.}
\graphicspath{{./}{figure/}}
\newcommand{\R}[1]{\textcolor{red}{#1}}
\newcommand{\B}[1]{\textcolor{blue}{#1}}
\usepackage{hyperref}
\usepackage{natbib}

\begin{document}
\title{Rapid Growth of Intermediate-Mass Black Holes through Disk-induced Stellar Disruptions}

\author[0000-0001-5019-4729]{Mengye Wang}
\affiliation{National Gravitation Laboratory, Hubei Key Laboratory of Gravitation and Quantum Physics, School of Physics, Huazhong University of Science and Technology, Luoyu Road 1037, Wuhan, China}
\affiliation{Department of Astronomy, School of Physics, Huazhong University of Science and Technology, Luoyu Road 1037, Wuhan, China}

\author[0000-0001-7192-4874]{Yiqiu Ma}
\altaffiliation{myqphy@hust.edu.cn}
\affiliation{National Gravitation Laboratory, Hubei Key Laboratory of Gravitation and Quantum Physics, School of Physics, Huazhong University of Science and Technology, Luoyu Road 1037, Wuhan, China}
\affiliation{Department of Astronomy, School of Physics, Huazhong University of Science and Technology, Luoyu Road 1037, Wuhan, China}

\author[0000-0003-4773-4987]{Qingwen Wu}
\altaffiliation{qwwu@hust.edu.cn} 
\affiliation{Department of Astronomy, School of Physics, Huazhong University of Science and Technology, Luoyu Road 1037, Wuhan, China}

\author[0000-0002-4966-7450]{Boyuan Liu}
\affiliation{Institut für Theoretische Astrophysik, Zentrum für Astronomie, Universität Heidelberg, Albert Ueberle Str. 2, D-69120 Heidelberg, Germany}

\author[0009-0007-3021-6266]{Zijian Wang}
\affiliation{National Gravitation Laboratory, Hubei Key Laboratory of Gravitation and Quantum Physics, School of Physics, Huazhong University of Science and Technology, Luoyu Road 1037, Wuhan, China}

\author[0009-0006-2353-666X]{Zhili Wang}
\affiliation{National Gravitation Laboratory, Hubei Key Laboratory of Gravitation and Quantum Physics, School of Physics, Huazhong University of Science and Technology, Luoyu Road 1037, Wuhan, China}

\begin{abstract}
Dense nuclear star clusters provide unique environments for studying the dynamical interactions between stars and massive black holes. When an accretion disk is present, dissipative star--disk interactions can capture surrounding stars, drive their inward migration, and ultimately lead to disk-induced tidal disruption events\,(dTDEs). The long-term feeding rate from this process, however, cannot be inferred from single-orbit migration estimates alone, as it depends on the coupled evolution of disk capture, collisional relaxation, stellar depletion and replenishment, and physical mergers within the star cluster.
In this work, we use high-performance direct $N$-body simulations combined with analytic prescriptions for star--disk interactions to follow this coupled evolution for intermediate-mass black holes\,(IMBHs) with accretion disks embedded in dense stellar clusters. The simulations track the formation of the stellar cusp, the capture of stars by repeated disk crossings, their subsequent orbital damping and migration, and their eventual consumption by the central IMBH. We find that dTDEs can sustain stellar mass supply rates of $\sim10^{-3}\,M_\odot \,\mathrm{yr}^{-1}$, which exceeds the Eddington-limited gas accretion rate for IMBHs with $M_\bullet<10^5\,M_\odot$. These results identify dTDEs as an efficient stellar feeding channel for IMBHs in gas-rich dense stellar systems. As one possible application, this mechanism may help transform $\sim10^3\,M_\odot$ IMBHs into more massive black-hole seeds, provided that compact stellar clusters and accretion disks persist for $>30$ Myr.
\end{abstract}


\section{INTRODUCTION} \label{sec:intro}
Dense star clusters in galactic nuclei provide unique laboratories for studying the complex dynamical interactions between stars and supermassive black holes\,(SMBHs). 
These systems are not only crucial for understanding the evolution of galactic nuclei, but also powerful probes of SMBH demographics, strong-field gravity, and accretion physics\,\citep[e.g.,][]{Kormendy2013ARA&A,Neumayer2020A&ARv}. 
In such environments, two-body relaxation can scatter stars onto low-angular-momentum orbits, driving a variety of astrophysical phenomena, including extreme mass-ratio inspirals\,(EMRIs) and tidal disruption events\,(TDEs)\,\citep[e.g.,][]{Merritt2013,Alexander2005PhR,Pau2022hgwa}. 
Classical TDEs in quiescent galactic nuclei have been extensively studied both theoretically and observationally in recent decades\,\citep[e.g.,][]{Rees1988Natur,Stone2020SSRv,Gezari2021ARA&A}. 
They are predominantly produced by stars approaching the SMBH on highly eccentric, nearly parabolic orbits as a result of two-body relaxation.
The presence of an accretion disk, however, opens up a fundamentally different route for feeding stars to the central black hole.
Stars repeatedly crossing the disk experience aerodynamic drag and gas dynamical friction\,\citep{Adachi1976PThPh,Ostriker1999ApJ}, which efficiently remove their orbital energy and angular momentum and gradually capture them onto the disk. 
Once embedded, they migrate inward through Type-I/II migration processes\,\citep[e.g.,][]{Goldreich1978,Goldreich1980}, eventually reaching the tidal disruption radius and giving rise to disk-induced tidal disruption events\,(dTDEs). 
Unlike classical TDEs, dTDEs are expected to occur on much lower-eccentricity orbits, potentially leading to distinct observational signatures. 
More importantly, for black hole growth, star--disk interactions can act as a dissipative conveyor of stellar material: the disk captures stars from a three-dimensional cluster, damps their orbits, and delivers them inward on a migration timescale.

This mechanism is especially relevant for intermediate-mass black holes\,(IMBHs). 
For $M_\bullet\lesssim10^5\,M_\odot$, the dTDE mass supply rate can exceed the Eddington-limited gas accretion rate, and may therefore dominate the early mass budget of the black hole.
The problem is not simply whether an individual star can be captured by the disk; that can be estimated from local drag and migration timescales. 
The central question is whether a realistic collisional stellar cluster can maintain a continuous supply of stars to the disk after the innermost stars have been consumed. 
The answer depends on several coupled processes: the formation of a stellar cusp around the IMBH, two-body relaxation that refills the disk-crossing population, stochastic few-body encounters, mass segregation and stellar evolution, and physical collisions among stars that captured by the disk. 
These effects determine both the dTDE rate and the mass spectrum of the disrupted stars.

Previous studies have proposed dTDEs as a possible mechanism for triggering enhanced nuclear activity\,\citep{Lei2026ApJ}, explaining the changing-look behavior in some AGNs\,\citep{wangyh2024arXiv,wangyh2024ApJ}, and accelerating SMBH mass growth\,\citep[e.g.,][]{Just2012ApJ,Kennedy2016MNRAS}. 
These studies have mainly focused on SMBHs in the $10^6$--$10^9\,M_\odot$ range. 
In particular, \citet{Kennedy2016MNRAS} found that dTDEs contribute negligibly to SMBH growth in this mass range, although their simulations were limited by the large particle numbers required for SMBH systems and therefore did not constitute a star-by-star treatment, nor did they account for physical collisions among stars captured into the accretion disk.
For IMBHs below $10^5\,M_\odot$, however, dTDEs could be crucial for accelerating their early growth. 
In this work, we investigate the mass growth of IMBHs driven by dTDEs using high-precision direct star-by-star $N$-body simulations with analytic prescriptions for star--disk interactions.
This approach is essential because the dTDE rate emerges from the live collective dynamics of the stellar system: stars are scattered onto new orbits via multi-body encounters, captured by the disk, and may merge, resulting in a population that is continuously consumed and replenished.
By modeling IMBHs embedded in dense stellar clusters with accretion disks, we track the entire sequence from cusp formation and disk capture to inward migration and tidal disruption, enabling us to quantify the stellar feeding rate of IMBHs.

Although our simulations are not designed to model a specific cosmological environment, the measured feeding rates have a natural implication for black hole seed growth. 
Wide-field optical and near-infrared surveys, including the \textit{Sloan Digital Sky Survey}, the DESI Legacy Imaging Surveys, and the \textit{James Webb Space Telescope}\,(JWST), have revealed SMBHs with masses of $\sim10^{6}$--$10^{10}\,M_\odot$ at $z\gtrsim6$\,\citep[e.g.,][]{Inayoshi2020ARA&A,Fan2023ARA&A,Maiolino2024A&A,Furtak2024Natur}. 
Simple Eddington-limited growth arguments imply that their progenitor black hole seeds must have already reached masses of $M_\bullet\gtrsim10^5 M_\odot$ by $z\sim15$\,\citetext{see Figure~31 in \citealp{Alexander2025NewAR}, or Figure~6 in \citealp{Taylor2025ApJ}}.
However, it remains unclear whether such heavy seeds can form efficiently at Cosmic Dawn, since their formation mechanisms and conditions are still debated\,\citep{Omukai2008ApJ,Latif2016ApJ,Chon2020MNRAS,Schleicher2022MNRAS,Kimura2025ApJ,Cenci2025MNRAS,Brennan2025OJAp,Chon2026arXiv}, whereas the formation of IMBHs with masses of $10^3$--$10^4\,M_\odot$ is considered more plausible\,\citep[e.g.,][]{Zwart2004Natur,Chon2020MNRAS,Rantala2024MNRAS,Fujii2024Sci,Vergara2025arXiv}. Once these IMBH seeds have formed, they can grow further through tidal disruption of stars and direct stellar collisions in the absence of an accretion disk\,\citep[e.g.,][]{Stone2017MNRAS,Rose2022ApJ,Gonzalez2025ApJ}, including the tidal disruption of Population III stars\,\citep[][]{WZJ2025ApJ}. In this work, we investigate whether an accretion disk embedded in a live dense cluster can continuously capture and deliver cluster stars to the IMBH. As an application, we use our simulated dTDE rates to illustrate how such disk-driven stellar feeding could enable the IMBH to grow rapidly toward $\sim10^5\,M_\odot$, provided that compact, gas-rich clusters are present.

The structure of this paper is as follows. Section~\ref{sec:model} describes the numerical strategy, initial conditions, and implementation of star--disk interactions. Section~\ref{sec:results} presents the evolution of the stellar cusp, the capture and migration of stars in the disk, the resulting dTDE rates, and an illustrative implication for IMBH seed growth. Finally, we summarize our work and discuss its limitations in Section~\ref{sec:discussion}.

\section{METHOD} \label{sec:model}
Our numerical model is designed to describe a dynamical system from which the dTDE feeding rate can be derived: a live collisional stellar cluster surrounds an IMBH, while an accretion disk provides interactions that capture and migrate stars. 
The cluster dynamics are evolved with the high-performance direct $N$-body code {\tt PETAR}\,\citep{WangLong2020MNRAS}, which is built on the {\tt FDPS} framework\,\citep[][]{Iwasawa2016PASJ,Iwasawa2020PASJ}, adopts a particle--tree particle--particle algorithm\,\citep{Oshino2011PASJ}, and incorporates the {\tt SDAR} method to accurately resolve close encounters and tight binaries\,\citep{WangLong2020MNRAS493}. The gas effects are included through analytic prescriptions for star--disk interactions. This hybrid approach allows us to keep the gravitational scattering, cusp formation, stellar mergers, and relaxation-driven replenishment fully self-consistent, while treating the unresolved gas response through physically motivated drag and migration formulae. In the following, we first describe the stellar cluster and IMBH initialization, then the disk forces, and finally the simulation suite used to measure the dTDE feeding rate.

\subsection{Initial conditions} \label{sec:initial}
The initial cluster specifies the stellar reservoir that can refill disk-crossing orbits after the inner region is consumed. 
We generate this reservoir using the {\tt MCLUSTER} code\,\citep{Kupper2011MNRAS}, adopting a total stellar mass of $M_{\rm cl}$ and a Plummer density profile\,\citep{Plummer1911MNRAS},
\begin{equation}
\rho(r) = \frac{3M_{\rm cl}}{4\pi a_{\rm P}^{3}}
\left(1+\frac{r^{2}}{a_{\rm P}^{2}}\right)^{-5/2},
\end{equation}
where $a_{\rm P}$ is the Plummer scale radius. The corresponding half-mass radius is given by $R_{\rm hm} \simeq 1.3\,a_{\rm P}$. The cluster is initialized in virial equilibrium without primordial mass segregation, and no primordial binaries are included in the initial conditions. Stellar masses are sampled from the initial mass function\,(IMF) of \citet{Kroupa2001MNRAS}, which is implemented as a broken power law. The IMF has a slope $\alpha_1 = -1.3$ for stellar masses in the range $0.08 \le m_\star/M_{\odot} \le 0.5$, with a Salpeter slope $\alpha_2 = -2.3$ for $0.5 < m_\star/M_{\odot} \le 150$. Subsequent stellar evolution is modeled using the {\tt SSE} and {\tt BSE} packages\,\citep{Hurley2000MNRAS,Hurley2002MNRAS}, which account for single and binary stellar evolution, respectively. A solar metallicity is adopted throughout the simulations.

To incorporate the central IMBH, we follow the procedure described in Section 3.2 of \citet{Lee2025ApJ}. Initially, the IMBH is placed at the center of the star cluster. To avoid numerical singularities associated with the sudden introduction of such a deep gravitational potential, the central IMBH is initially treated as an external potential rather than as a live particle. Its mass is raised adiabatically from zero to $M_\bullet$ over a timescale of $2\,\mathrm{Myr}$. This gradual growth allows the stellar system to dynamically adjust to the emerging central potential. After $2\,\mathrm{Myr}$, once the IMBH mass has reached its target value $M_\bullet$, the external potential is replaced by a live $N$-body particle of the same mass. From this point onward, the IMBH fully participates in gravitational interactions and evolves self-consistently with the surrounding stellar system. Note that stellar evolution is switched on from the very beginning of the simulation, including the $2\,\mathrm{Myr}$ adiabatic IMBH growth phase.

\begin{figure*}
\centering
\includegraphics[scale=0.45]{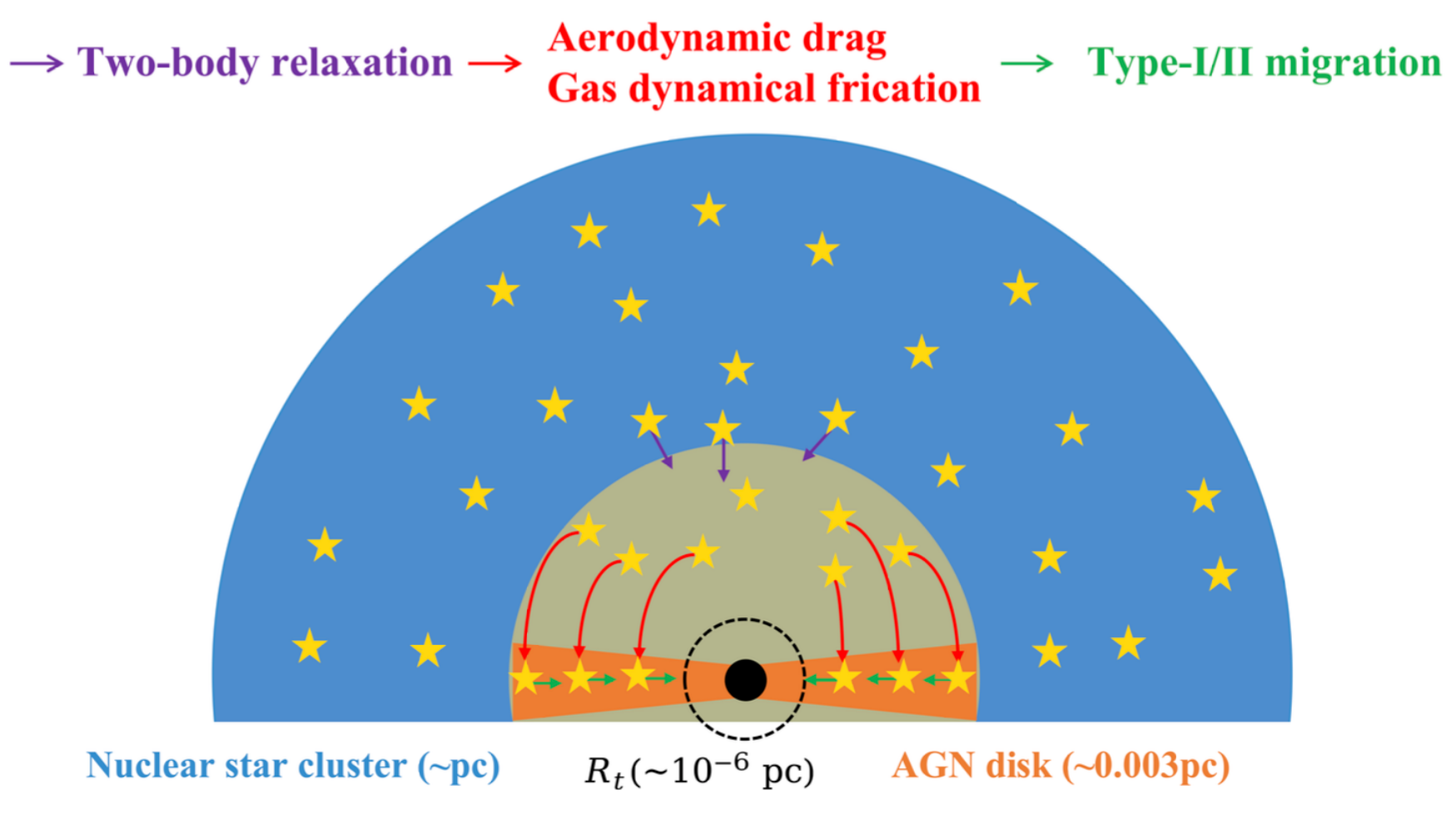}
\caption{Schematic illustration of the dynamical evolution of a star cluster under the influence of an IMBH accretion disk. Stars passing through the disk are gradually captured onto the disk plane via aerodynamic drag or gaseous dynamical friction. Subsequently, they migrate inward through Type~I/II migration, eventually crossing the tidal disruption radius and being consumed by the IMBH. As stars in the inner region are consumed, stars from the outer cluster are continuously replenished inward via two-body relaxation, establishing a steady supply of stars to the central IMBH. }
\label{fig:cartoon}
\end{figure*}

\subsection{Star-Disk Interactions}

When the central IMBH is in an active accretion state, its accretion disk can strongly influence the dynamical evolution of the surrounding nuclear star cluster. 
The role of the disk in our calculation is to provide the dissipative step that ordinary stellar dynamics lacks: it converts disk-crossing orbits into embedded, migrating orbits that can be consumed by the IMBH. 
In this subsection, we summarize the analytical prescriptions for star--disk interactions and describe their implementation as external forces in the {\tt PETAR} code for $N$-body simulations.


Star--disk interactions have been extensively investigated in previous studies\,\citep[e.g.,][]{Cresswell2008A&A,McKernan2012,Macleod2020ApJ,Panzhen2021PhysRevD,Wang2023MNRAS,wmy2024ApJ,Rowan2025MNRAS,Whitehead2025MNRAS}. Here we summarize the key prescriptions adopted in this work:

(1) For stars with high inclinations and eccentricities relative to the disk midplane, the interaction during each disk crossing is dominated by two related processes: aerodynamic drag\,\citep{Adachi1976PThPh} and gas dynamical friction\,\citep{Ostriker1999ApJ}. These can be unified as:
\begin{align} \label{eq:GDF}
  \vec{a}_{\rm drag} &= \rho x^2 f(x) \pi R_{\rm drag}^2 \frac{\vec{v}-\vec{v}_{\rm K}}{|\vec{v}-\vec{v}_{\rm K}|},\\ 
  f(x) &= 
  \begin{cases}
    \dfrac{1}{2} \ln \left( \dfrac{1+x}{1-x}\right) -x, & 0< x \leq 1, \\[1.2ex]
    \dfrac{1}{2}\ln \left(x^2-1\right) + 3.1, & x >1,
  \end{cases}
\end{align}
where $x \equiv |\vec{v}-\vec{v}_{\rm K}|/c_{\rm s}$ is the Mach number of the star's motion relative to the background gas, and $\vec{v}_{\rm K}$ is the local Keplerian velocity. The effective cross-section is determined by the drag radius:
\begin{equation}
R_{\rm drag} = \max\left(r_\star,\, \frac{2Gm_\star}{x^2}\right),
\end{equation}
where $r_\star$ and $m_\star$ are the stellar radius and mass. For low-mass stars, $R_{\rm drag}\sim r_\star$\,(aerodynamic drag); for massive stars or compact objects, gravitational focusing dominates with $R_{\rm drag}\sim 2Gm_\star/(\vec{v}-\vec{v}_{\rm K})^2$\,(gas dynamical friction).

(2) For stellar objects with low orbital inclinations and eccentricities, or those embedded within the disk, star--disk interactions are predominantly mediated by density waves, leading to Type I/II migration\,\citep[][]{Goldreich1979ApJ,Goldreich1980,Tanaka2002ApJ,Tanaka2004ApJ,LYP2024ApJ}. Through hydrodynamical simulations, \citet{Cresswell2008A&A} provided empirical fitting formulae for the damping timescales of semi-major axis $a$, eccentricity $e$, and inclination $i$ in the Type I regime:
\begin{equation}   \label{eq:tau_mig}
\begin{split}
 & \frac{\tau_{\mathrm{mig,I}}}{\tau_{\mathrm{wave}}} = \frac{P(e)}{h^2} \left[1+\frac{0.070\eta_i+0.085\eta_i^4-0.080\eta_e\eta_i^2}{|P(e)|} \right],\\
 &\frac{\tau_{\mathrm{i}}}{\tau_{\mathrm{wave}}}=1.84(1-0.30\eta_i^2+0.24\eta_i^3 +0.14\eta_e^2\eta_i), \\
 & \frac{\tau_{\mathrm{e}}}{\tau_{\mathrm{wave}}}=1.28(1-0.14\eta_e^2+0.06\eta_e^3+0.18\eta_e\eta_i^2),
\end{split}
\end{equation}
where $\eta_i\equiv i/h,\eta_e\equiv e/h$, $h\equiv H/R$ is the disk aspect ratio and 
\begin{equation}
  \begin{split}
    & P(e)=\frac{1+\left(\eta_e/2.25\right)^{1.2}+\left(\eta_e/2.84\right)^6}{1-\left(\eta_e/2.02\right)^4},\\
    & \tau_{\rm wave} = \frac{M_\bullet}{m_\star}\frac{M_\bullet}{\Sigma a^2} \frac{h^4}{\Omega}.
  \end{split}
\end{equation}
Here $\beta$ is the surface density gradient\,($\Sigma\propto r^{-\beta}$), $\Omega$ is the orbital frequency. 

When the embedded object is sufficiently massive to open a gap in the disk\,\citep{Goldreich1980, lin1986}, migration transitions to the Type II regime. The gap-opening criterion is\,\citep{Crida2006Icar}:
\begin{equation}
\frac{3H}{4R_{\mathrm{H}}} + \frac{50\nu}{q r \Omega_\star} \lesssim 1 \quad \mathrm{and} \quad R_{\mathrm{H}} \gtrsim H,
\label{eq:gap}
\end{equation}
where $R_{\mathrm{H}} = r\,(m_\star/3M_\bullet)^{1/3}$ is the Hill radius, $H$ is the disk scale height, $\nu$ is the viscosity, and $q=m_\star/M_\bullet$ is the mass ratio. Once a gap opens, the object migrates on the viscous timescale:
\begin{equation}
\tau_{\mathrm{mig,II}} \sim \frac{r^2 \Omega}{\Gamma_{\rm mig, II}} \sim \frac{M_\star}{2\pi r \Sigma |v_{\mathrm{gas},r}|},
\end{equation}
where $v_{\mathrm{gas},r}\sim -\dot{M}/(2\pi r \Sigma)$ is the radial gas velocity. This timescale is typically much longer than Type I migration.

The damping effects from density waves can be implemented as additional acceleration terms in the $N$-body simulations. Following the timescales derived above, we apply the following accelerations to describe the star--disk interactions:
\begin{equation} \label{eq:a_dw}
    \vec{a}_{\rm DW} = -\frac{\vec{v}}{\tau_{\mathrm{mig}}}  -2\frac{(\vec{v}\cdot\vec{r})\vec{r}}{r^2 \tau_{\mathrm{e}}} -\frac{v_z}{\tau_{\mathrm{i}}} \hat{{z}}
\end{equation}
where $\boldsymbol{v}$ and $\boldsymbol{r}$ are the velocity and position vectors of the stellar-mass objects relative to the central IMBH, $v_z$ is the vertical velocity component, $\hat{{z}}$ is the unit vector perpendicular to the disk midplane.

In general, the drag force (equation~\eqref{eq:GDF}) and the torque from density waves (equation~\eqref{eq:a_dw}) are implemented as external forces in the {\tt PETAR} $N$-body simulations to model the capture of stars into the accretion disk and their subsequent inward migration. Both forces are applied simultaneously to all stars: for stars on high-inclination orbits, equation~\eqref{eq:GDF} dominates the interaction, while equation~\eqref{eq:a_dw} becomes increasingly important as a star settles onto the disk plane.

\subsection{Simulation setup}
We perform a suite of simulations designed to measure how the dTDE feeding rate depends on the IMBH mass while keeping the surrounding stellar reservoir fixed. 
The three central IMBH masses considered are $M_\bullet = 3\times 10^3$, $10^4$, and $3\times 10^4\,M_\odot$. 
The initial star cluster, generated as described in Section~\ref{sec:initial}, has a total mass of $M_{\rm cl}=2\times 10^5\,M_\odot$, a half-mass radius of $R_{\rm hm}=0.6\,\rm pc$, and contains approximately $3.7\times 10^5$ particles. 
This configuration is compact enough for two-body relaxation to repopulate the central region on Myr timescales, while still sufficiently extended to host a live stellar reservoir whose depletion and refilling are governed self-consistently by the simulation rather than imposed externally.

For the accretion disk, we adopt the SG disk model\,\citep{Sirko2003} with a viscosity parameter $\alpha = 0.1$. 
The gas accretion rate onto the central IMBH is set to the Eddington rate, $\dot{M}_{\rm gas} = \dot{M}_{\rm Edd}$, and the outer boundary of the disk is taken as $R_{\rm out}=R_{\rm sg}\simeq 0.003\,\mathrm{pc}$, where $R_{\rm sg}$ is the characteristic transition radius of the SG model. 
Within $R_{\rm sg}$, the disk mass is less than $\sim5\%$ of the IMBH mass, so its gravitational influence on stellar dynamics is negligible. 
However, the region beyond $R_{\rm sg}$ consists of an outer star-forming disk regulated at $Q\sim1$, whose extent is uncertain and depends on the balance between stellar feedback and radiative cooling. 
This extended gaseous component could have a considerable mass, and dynamical friction with stars at larger radii may accelerate stellar replenishment into the inner disk. 
However, extending the disk boundary into this region also significantly increases the number of stars captured. 
In our tests with $R_{\rm out}=0.1\,\rm pc$, the enhanced stellar density leads to frequent collisions and mergers, forming very massive stars ($>100\,M_\odot$) that can perturb the disk structure around a $\sim10^4\,M_\odot$ IMBH, challenging the assumption of a steady-state disk. 
To avoid these uncertainties, we restrict our fiducial model to the inner standard disk with $R_{\rm out}=R_{\rm sg}$. 
The impact of a moderately extended disk ($R_{\rm out}=3R_{\rm sg}\sim0.01\,\mathrm{pc}$) is examined in Section~\ref{sec:discussion}.

Figure~\ref{fig:cartoon} presents a schematic illustration of our model. 
The simulation is divided into two stages that separate the formation of the stellar background from the subsequent disk-driven feeding process. 
First, the central IMBH is introduced adiabatically over $2\,\mathrm{Myr}$ following the procedure described in Section~\ref{sec:initial}, allowing the cluster to respond to the central potential and form the stellar distribution from which the disk will later draw stars. 
After the IMBH reaches its target mass, it is converted into a live particle and the star--disk interaction forces are activated. 
Stars interacting with the disk are then gradually captured onto near-circular, coplanar orbits through the damping forces described above, and once captured, they migrate inward via Type~I/II migration, eventually approaching the tidal disruption radius of the central IMBH. 
The second stage, during which star--disk interactions are active, is evolved for only $\sim 0.3$--$0.5\,\mathrm{Myr}$. This short duration is dictated by the computational cost of the direct $N$-body integration: once a large number of stars are captured onto the disk, they and the central IMBH form a compact few-body cluster with short orbital timescales, which forces the use of a very small timestep and makes the integration extremely slow.
By evolving both stages in the same direct $N$-body calculation, we can measure whether the dTDE rate is sustained by relaxation-driven replenishment rather than by the initially available inner stars alone.


In our simulations, a star is considered accreted once its semi-major axis $a$ becomes smaller than $R_{\rm dTDE}$. We adopt $R_{\rm dTDE}=10^{-4}\,\mathrm{pc}$ as our fiducial value, which is much larger than the actual physical tidal disruption radius $R_{\rm t}$\,($\sim 10^{-6}\, \rm pc $). This choice is primarily motivated by the computational cost associated with resolving the short orbital timescales near the central IMBH. For a representative IMBH mass of $M_\bullet=10^4\,M_\odot$, the Keplerian orbital period at $R_{\rm dTDE}=10^{-4}\,\mathrm{pc}$ is approximately $P_{\rm orb}\simeq1\,\mathrm{yr}$. Since our $N$-body integration requires a timestep smaller than $P_{\rm orb}/32$, resolving a significantly smaller accretion radius would substantially increase the computational cost. Therefore, we use this $R_{\rm dTDE}$ as a proxy for disruption, and this approximation can be justified by the fact that captured stars follow well-defined, deterministic migration trajectories\citep[][]{Just2012ApJ,Kennedy2016MNRAS}. Even for stars that are not yet fully embedded in the disk, once they cross $R_{\rm dTDE}$, dynamical friction from the accretion disk will also drag them inward and lead to tidal disruption by the IMBH within $\sim 10^4$ yr\,\citep[e.g.,][]{wangyh2024ApJ}. For the same reason, extending our simulations to more massive IMBHs remains computationally challenging.
In Section~\ref{sec:discussion}, we also perform a controlled run with $R_{\rm dTDE}=5\times10^{-5}\,\mathrm{pc}$ to verify that our results are insensitive to this choice.

\begin{figure}[htbp]
\centering
\includegraphics[scale=0.45]{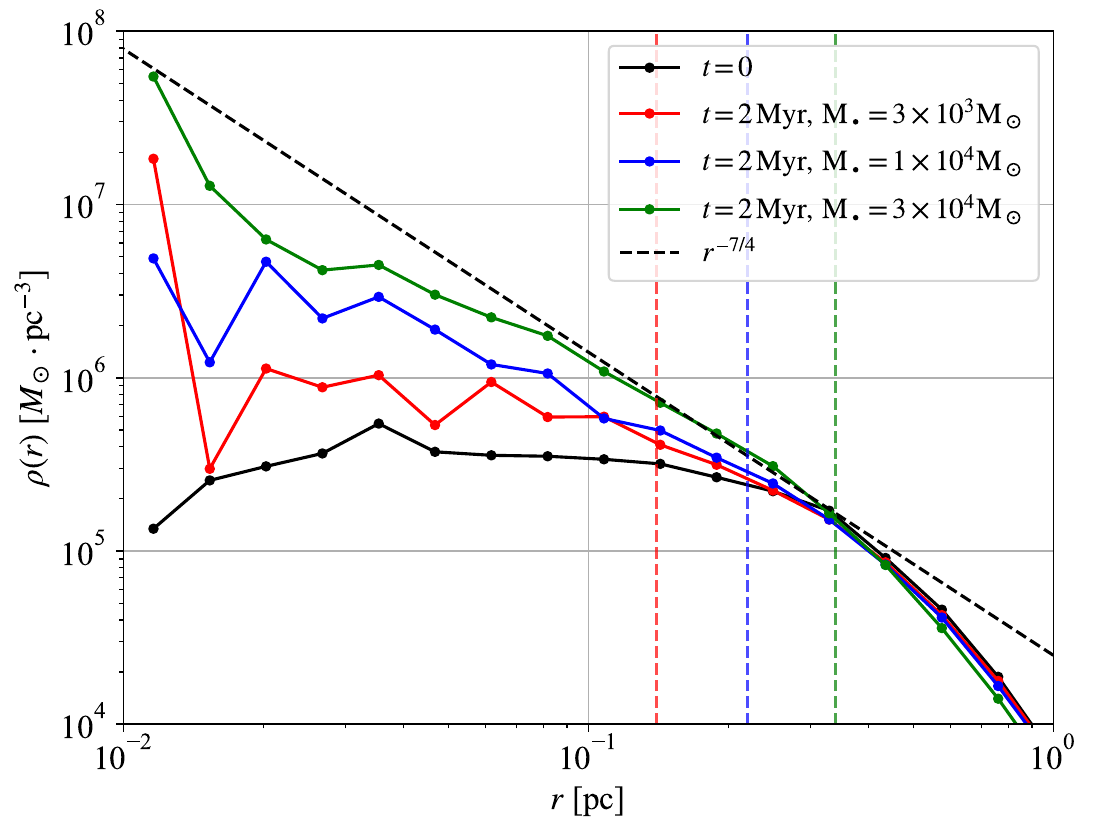}
\caption{Stellar density profiles at $t=2\,\mathrm{Myr}$, after the adiabatic growth of the central IMBH and before the activation of disk forces. Different colors correspond to different IMBH masses: $M_\bullet = 3\times 10^3\,M_\odot$\,(red), $10^4\,M_\odot$\,(blue), and $3\times 10^4\,M_\odot$\,(green). The black dashed line indicates the canonical Bahcall--Wolf slope $\gamma=-1.75$ for reference. The vertical dashed lines mark the corresponding influence radii $r_{\rm h}$, defined by $M_\star(r < r_{\rm h}) = 2M_\bullet$\,\citep[see][]{Merritt2013book}, using the same color coding for the three IMBH masses.}
\label{fig:star_density}
\end{figure}

\section{Results} \label{sec:results}

\begin{figure*}
\centering
\includegraphics[scale=0.54]{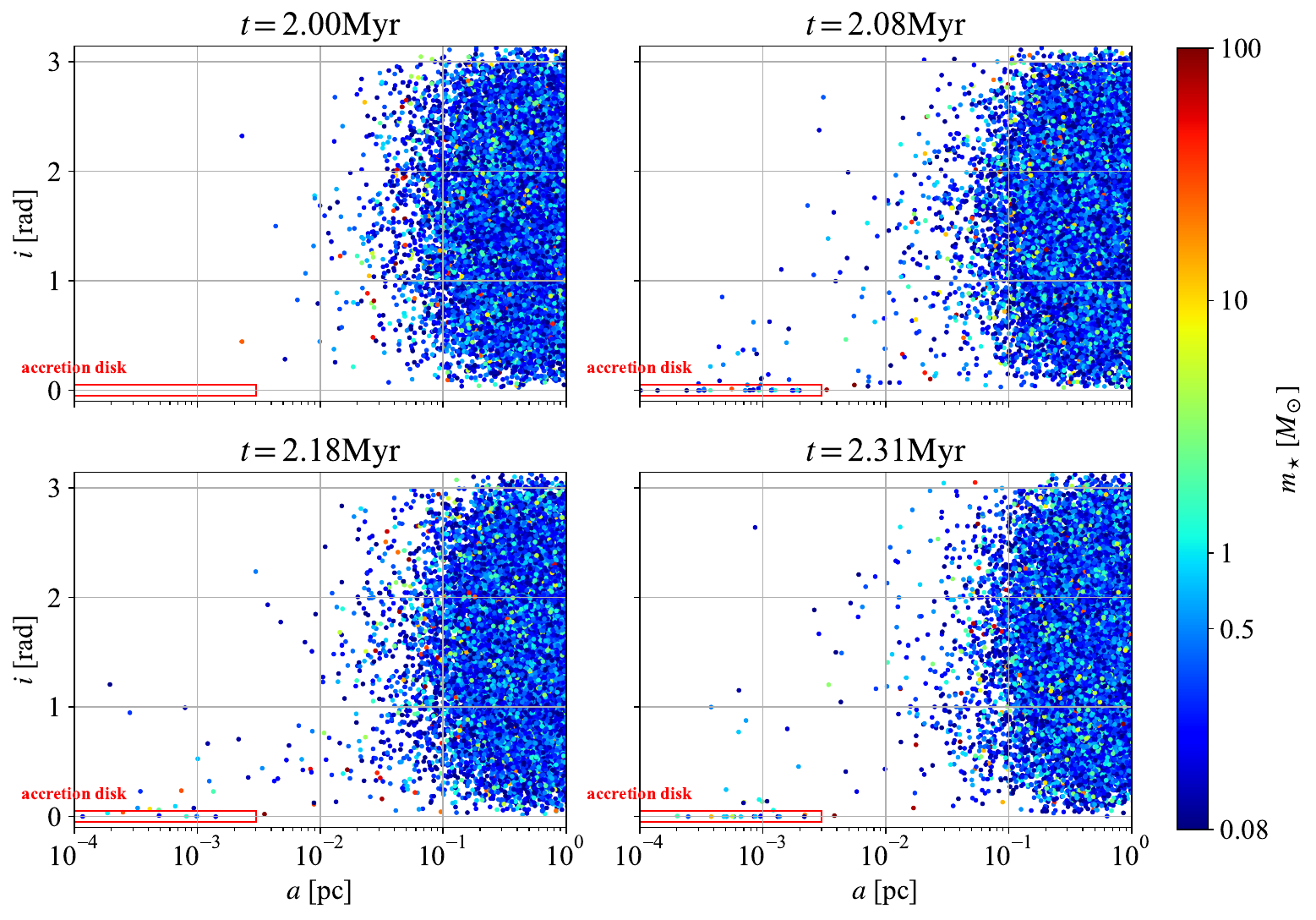}
\caption{Evolution of stellar orbital inclination. The four panels show the distribution of semi-major axis $a$ versus orbital inclination $i$ at $t = 2$, $2.05$, $2.2$, and $2.32\,\mathrm{Myr}$, respectively, with the colorbar representing stellar mass.
At $t = 2\,\mathrm{Myr}$, the orbital inclinations are nearly uniformly distributed between $0$ and $\pi$, and the probability of finding stars with semi-major axes smaller than $0.001\,\mathrm{pc}$ is very low. After the star--disk interactions are turned on, stars are rapidly captured onto the accretion disk with nearly zero inclination and subsequently migrate inward.}
\label{fig:orbital_inclination}
\end{figure*}

\subsection{Stellar density profile before disk activation} \label{sec:density_profile}

Before activating star--disk interactions, we first examine the quasi-equilibrium stellar distribution established around the central IMBH following its adiabatic growth. As shown in Figure~\ref{fig:star_density}, we present the stellar density profiles at $t=2\,\mathrm{Myr}$, immediately after the IMBH has reached its final mass and prior to the inclusion of disk forces, for three different IMBH masses: $M_\bullet = 3\times 10^3\,M_\odot$\,(red), $10^4\,M_\odot$\,(blue), and $3\times 10^4\,M_\odot$\,(green). As described in Section~\ref{sec:initial}, the cluster is initially modeled with a Plummer profile characterized by a flat central core. During the $2\,\mathrm{Myr}$ adiabatic growth phase, the stellar system responds to the deepening potential, leading to the formation of a central density cusp within the IMBH’s sphere of influence. As shown in Figure~\ref{fig:star_density}, the inner regions\,($r \lesssim 0.1\,\mathrm{pc}$) become significantly steeper than the initial profile, with the break radius broadly consistent with the corresponding influence radius.

Theoretically, two-body relaxation predicts a power-law cusp $\rho \propto r^{-1.75}$ for a relaxed stellar system around a massive black hole\,\citep{Bahcall1976ApJ,Bahcall1977ApJ}. Our results broadly follow this expectation, particularly for higher black hole masses. For $M_\bullet = 3\times 10^3\,M_\odot$, the inner cusp remains relatively shallow due to the weaker gravitational influence and limited relaxation over $2\,\mathrm{Myr}$. For the more massive central IMBH\,($M_\bullet = 3\times 10^4\,M_\odot$), the stellar density profile becomes significantly steeper, approaching $\gamma \sim -1.75$, consistent with the canonical Bahcall--Wolf solution.

The two-body relaxation timescale at the influence radius of IMBH, $r_{\rm h}$, is approximately $10$ Myr, which exceeds the $2$ Myr duration of the adiabatic growth phase adopted in our simulations. 
While this choice is primarily dictated by computational constraints, it is unlikely to significantly affect our inferred dTDE rates. 
Once the disk torque is switched on, the stellar distribution within $R_{\rm out}$ becomes progressively more cuspy as stars are efficiently captured and transported inward. 
Within this region, the local two-body relaxation timescale is below $\sim 10^4$ yr, whereas the characteristic inward migration timescale of stars through the disk is slightly longer than $\sim 10^4$ yr. 
Provided that the stellar replenishment rate in the vicinity of the disk, which is driven primarily by two-body relaxation and may be further enhanced by dynamical friction with gas clouds, exceeds the rate at which stars are consumed through inward migration, the dTDE rate can approach a quasi-steady state. 
This is consistent with the approximately constant dTDE rates observed over the duration of our simulations.

\subsection{Orbital Inclination Decay Induced by Star--Disk Interactions} \label{sec:inclination_decay}

After introducing the star--disk interactions at $t=2\,\mathrm{Myr}$, the accretion disk quickly imprints a strong dynamical signature on the surrounding stellar population. Figure~\ref{fig:orbital_inclination} shows the distribution of stars in the semi-major axis versus orbital inclination plane at $t = 2.00$, $2.08$, $2.18$, and $2.31\,\mathrm{Myr}$. At the initial moment when the disk is activated\,($t=2\,\mathrm{Myr}$, top-left panel), the stellar orbits exhibit an isotropic inclination distribution, spanning uniformly from $0$ to $\pi$ with a clear symmetry about $\pi/2$. This reflects the relaxed, nearly spherical configuration established during the adiabatic growth phase of the central IMBH. In this stage, the probability of finding stars with semi-major axes smaller than $0.001\,\mathrm{pc}$ is very low.

Once the disk forces are turned on, a rapid transformation occurs in the inner region. At $t>2\,\mathrm{Myr}$, stars within the disk outer radius\,($R_{\rm out}\approx 0.003\,\mathrm{pc}$) are efficiently captured onto near-coplanar orbits, forming a dense, coplanar structure with inclination $i \sim 0$. This alignment is driven by repeated disk crossings, during which aerodynamic drag and gas dynamical friction\,(see equation~\eqref{eq:GDF}) efficiently dissipate the vertical component of stellar motion and progressively align the angular momentum vectors with the disk plane. In contrast, stars outside $R_{\rm out}$ remain largely unaffected and retain their initial isotropic inclination distribution. Meanwhile, stars embedded in the accretion disk migrate inward via density wave excitations and eventually reach the accretion radius $R_{\rm dTDE}$.

\begin{figure*}
\centering
\includegraphics[scale=0.47]{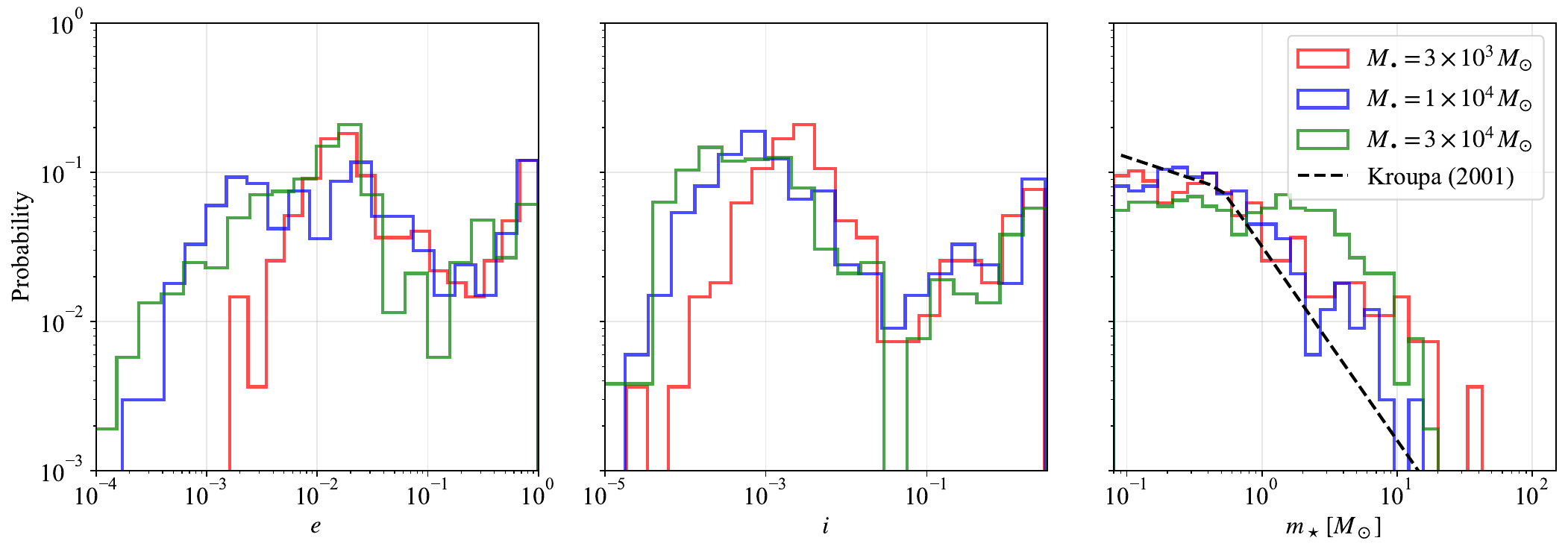}
\caption{Distributions of orbital parameters and masses for disrupted stars. The three panels from left to right show the eccentricity $e$, inclination $i$, and stellar mass $m_\star$, respectively. The vertical axis indicates the normalized number of disrupted stars per logarithmic bin. Red, blue, and green correspond to central IMBH masses of $3\times10^3$, $10^4$, and $3\times10^4\,M_\odot$, respectively. The black dashed line in the right panel shows the initial stellar IMF of the star cluster from \citet{Kroupa2001MNRAS}.}
\label{fig:parameters_TDE}
\end{figure*}

\begin{figure}[t]
\centering
\includegraphics[scale=0.42]{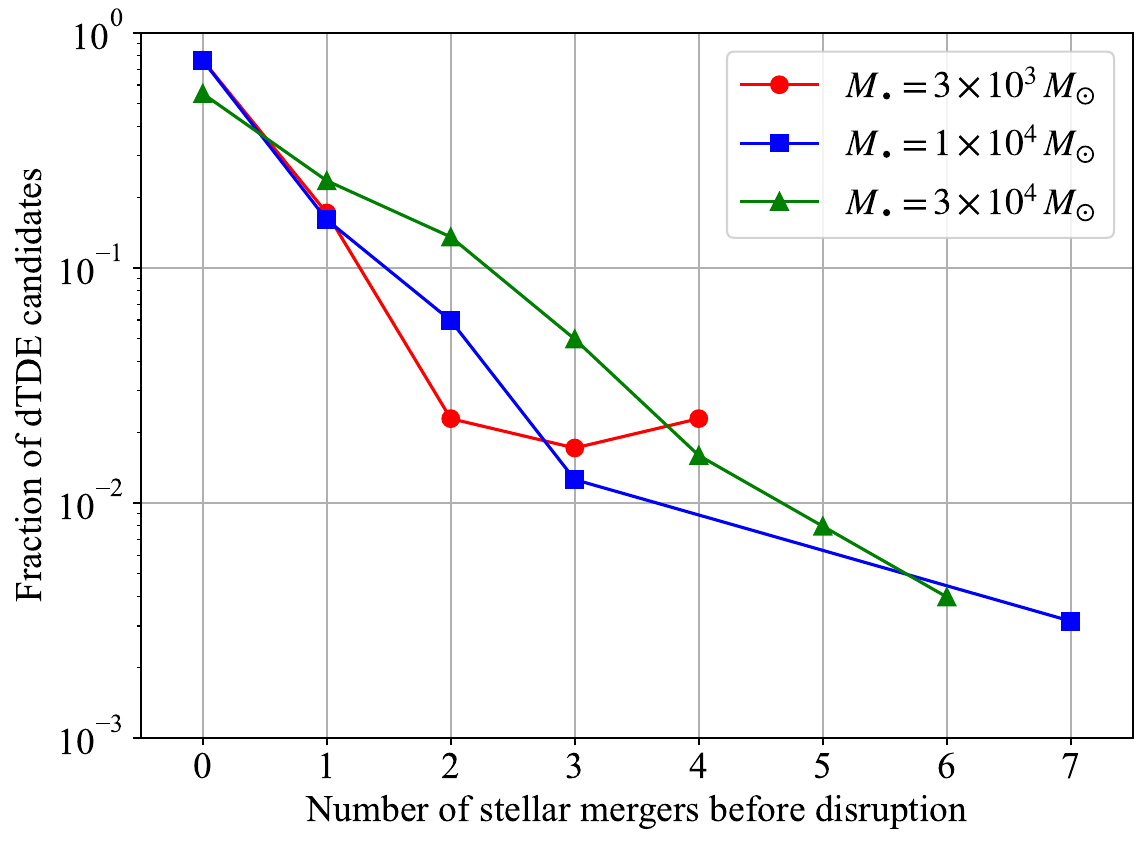}
\caption{Number of stellar collisions and mergers experienced by dTDE stars before disruption in the accretion disk, for different central IMBH masses.}
\label{fig:merger_statistics}
\end{figure}

\begin{figure}[t]
\centering
\includegraphics[scale=0.42]{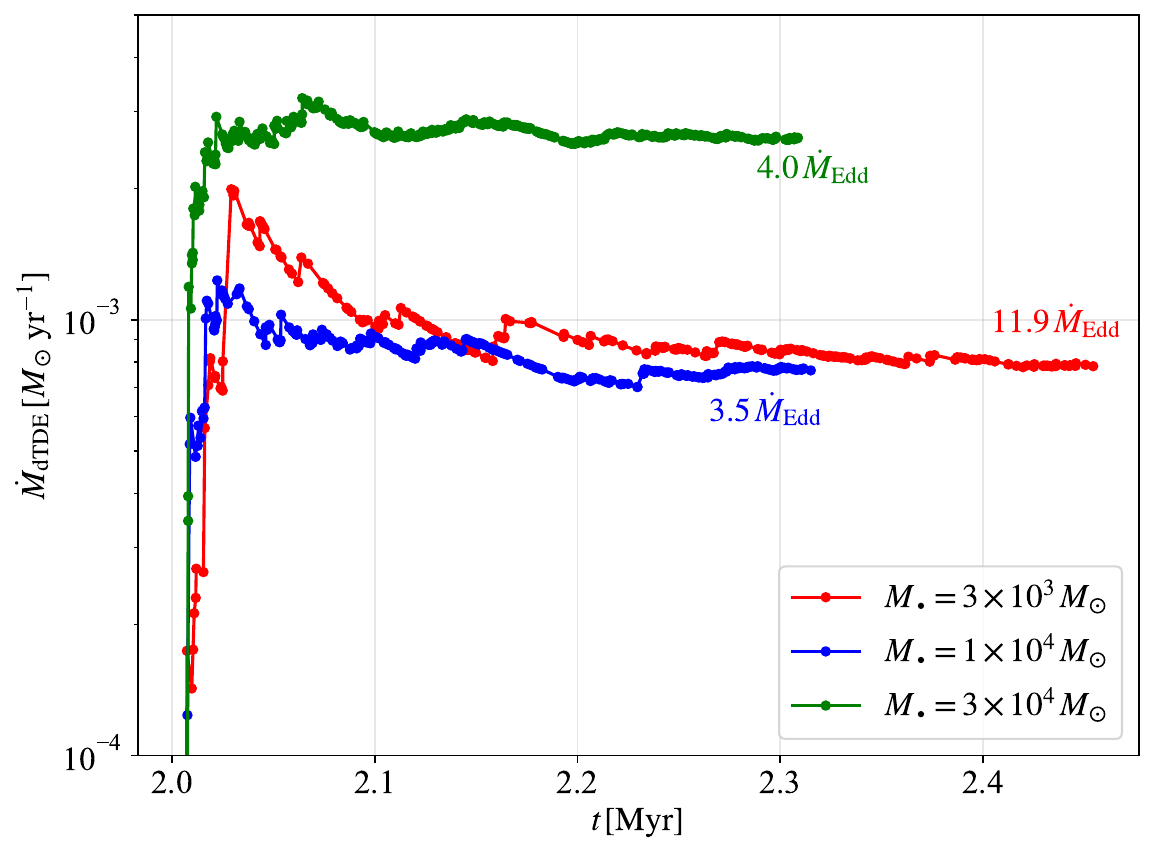}
\caption{Rate of dTDE as a function of time. The time-averaged stellar disruption rate is defined as $\dot{M}_{\rm dTDE}(t) \equiv M_{\rm tot}(t) / (t-2\,\rm Myr)$, where $M_{\rm tot}(t)$ is the cumulative mass of stars tidally disrupted by time $t$. Red, blue, and green curves correspond to central IMBH masses of $3\times10^3$, $10^4$, and $3\times10^4\,M_\odot$, respectively. The ratio of the disruption rate to the Eddington accretion rate at the end of the simulation is indicated for each black hole mass.}
\label{fig:TDE_rate}
\end{figure}

\subsection{Rate of disk-induced TDE} \label{sec:TDE}

In this subsection, we use the crossing of the inner accretion radius, $R_{\rm dTDE}=10^{-4}\,\mathrm{pc}$, as the operational definition of a dTDE candidate. 
This radius connects the local orbital outcome of star--disk interactions to the global feeding rate of the IMBH. 
The disruption rate is not set only by the instantaneous number of stars inside the disk. 
It is controlled by a sequence of linked processes: stars must first be supplied to disk-crossing orbits by the collisional cluster, then dissipatively captured or driven inward by the disk, and finally avoid being scattered or removed by physical stellar mergers before reaching $R_{\rm dTDE}$.

The orbital distributions of the disrupted stars therefore provide a useful diagnostic of the underlying feeding channels. 
As shown in Figure~\ref{fig:parameters_TDE}, the disrupted population is bimodal in both eccentricity and inclination. 
One component has low eccentricity\,($e\sim10^{-2}$) and very small inclination\,($i\sim10^{-3}\,\mathrm{rad}$), corresponding to stars that are first captured into the disk, circularized and aligned, and then transported inward by Type~I/II migration. 
This is the clean disk-mediated channel: dissipation is strong enough to erase the memory of the original three-dimensional orbit before disruption. 
The second component remains highly eccentric and often retrograde. 
These stars are also affected by the disk, but their semi-major axes decay faster than their inclination is damped, so they cross $R_{\rm dTDE}$ before becoming fully embedded. 
The coexistence of these two populations demonstrates that the dTDE sample is not described by a single circular migration track, but by the competition between energy loss, inclination damping, and scattering in the live cluster. We find that $\sim80\%$ of disrupted stars have orbital inclination below the disk aspect ratio ($i < h$) when their semi-major axis first drops below $R_{\rm dTDE}=10^{-4}\,\mathrm{pc}$. 
For the remaining $\sim20\%$ with $i > h$, they continue to interact with the accretion disk after entering $R_{\rm dTDE}$, which further damps their orbital inclination and eccentricity, eventually leading them into the true tidal disruption radius $R_{\rm t}$.

The same capture-and-migration physics also shapes the stellar masses delivered to the IMBH. 
The right panel of Figure~\ref{fig:parameters_TDE} shows that the disrupted stars are mildly top-heavy relative to the initial IMF. 
Massive stars are preferentially selected because their larger effective drag cross sections and faster migration make them easier to capture and deliver inward. 
The strength of this selection depends on the IMBH mass. 
For more massive IMBHs, the stellar-to-BH mass ratio is smaller, so massive stars can remain in the rapid Type~I migration regime and contribute efficiently to dTDEs. 
For lower-mass IMBHs, the larger mass ratio can move massive stars into slower Type~II migration, reducing their inward delivery. 
Once stars are compressed into the disk plane, physical collisions provide an additional growth channel before disruption. 
As shown in Figure~\ref{fig:merger_statistics}, the fraction of dTDE candidates that experience at least one merger is approximately 23.4\%, 23.6\%, and 44.8\% for $M_\bullet = 3\times10^3$, $10^4$, and $3\times10^4\,M_\odot$, respectively. 
The increasing merger fraction with IMBH mass is consistent with more efficient disk capture and migration, which raise the density of massive stars in the disk and increase their probability of colliding before they are consumed.

Combining these orbital and mass-selection effects gives the net stellar feeding rate shown in Figure~\ref{fig:TDE_rate}. 
For $M_\bullet = 3\times 10^3\,M_\odot$ and $10^4\,M_\odot$, the disruption rates remain at $\sim 10^{-3}\,M_\odot\,\mathrm{yr}^{-1}$, while the rate is $\sim 3\times 10^{-3}\,M_\odot\,\mathrm{yr}^{-1}$ for $M_\bullet = 3\times 10^4\,M_\odot$. 
These values are significantly higher than the classical relaxation-driven rate of TDEs\,($\sim 10^{-5}$--$10^{-4}\,M_\odot\, \mathrm{yr}^{-1}$). 
Although the curves fluctuate because individual disruptions are discrete events, they remain approximately steady over the simulated interval. 
This behavior indicates that the inner stellar population is not simply exhausted in an initial burst. 
Instead, the system approaches a quasi-steady state in which two-body relaxation continually feeds new stars toward the disk, while disk capture and migration remove them through dTDEs. 
The resulting dTDE-driven mass supply exceeds the Eddington-limited gas accretion rate for all three IMBH masses considered here, demonstrating that disk-induced stellar disruptions can dominate the early growth budget of low-mass IMBHs under the assumed cluster and disk conditions.

\subsection{Illustrative Implication for Seed Black Hole Growth}

\begin{figure}[t]
\centering
\includegraphics[scale=0.38]{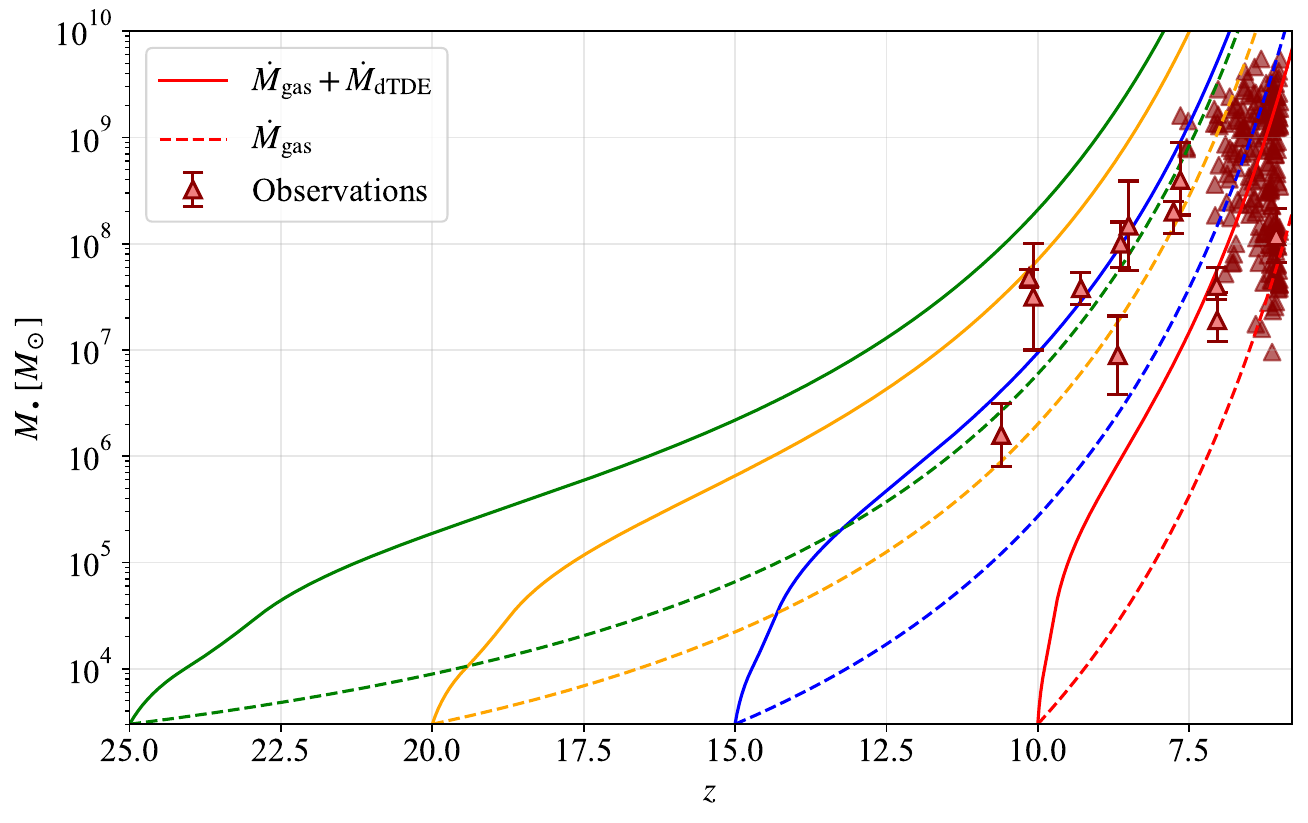}
\caption{Illustrative seed-growth histories based on the dTDE rates measured in our simulations. The dashed lines show growth via Eddington-limited gas accretion alone\,($\dot{M}_{\rm gas} = \dot{M}_{\rm Edd}$), while the solid lines show growth via the combined contribution of Eddington-limited gas accretion and dTDEs. Green, orange, blue, and red lines correspond to seed black holes of mass $3\times10^3\,M_\odot$ formed at redshifts $z = 25$, $20$, $15$, and $10$, respectively. The triangles indicate observed supermassive black holes at redshifts $z > 6$\,\citep[see][and references therein]{Taylor2025ApJ}. This comparison is intended as a conditional implication of the simulated feeding rates rather than a systematic cosmological model; a full study of seed black hole growth in cosmological environments will be presented in a companion paper.}
\label{fig:BH_growth}
\end{figure}

The simulations above measure the dTDE rate as a local stellar feeding process in gas-rich dense clusters. 
Here we use these rates to illustrate one possible implication for black hole seed growth, without assuming that our cluster model represents the full complexity of high-redshift galaxy formation. 
Figure~\ref{fig:BH_growth} should therefore be read as an illustrative implication of the simulated feeding rates. 
A systematic study of seed black hole growth within cosmological environments requires cosmological simulations and will be presented in a companion paper.

As discussed in the Introduction, the observed population of high-redshift quasars requires the existence of black hole seeds with masses of $M_\bullet\gtrsim10^5\,M_\odot$ by $z\sim15$. However, the formation of such massive seeds remains challenging. The light-seed scenario, in which stellar-mass black holes form from Population~III stars at $z\sim20$--$30$\,\citep{Haiman2001ApJ,Madau2001ApJ}, generally requires prolonged super-Eddington accretion to reach the observed SMBH masses, which is difficult to sustain due to radiative and mechanical feedback from the accreting black hole\,\citep[e.g.,][]{Sassano2023MNRAS,Shi2023MNRAS,wu2025ApJ,Kiyuna2026MNRAS}. The heavy-seed scenario through direct collapse of primordial gas clouds can produce black holes with masses of $\sim10^{4}$--$10^{6}\,M_\odot$\,\citep[e.g.,][]{Oh2002ApJ,Bromm2003ApJ,Begelman2006MNRAS}. Related TDE-assisted growth channels, including tidal disruptions of Population~III stars by IMBH seeds, have also been proposed to boost early black hole growth\,\citep{WZJ2025ApJ}. However, these scenarios are highly sensitive to fragmentation induced by trace amounts of dust, metals, or molecular hydrogen\,\citep{Omukai2008ApJ,Latif2016ApJ,Chon2020MNRAS}. Such fragmentation may favor the formation of multiple massive stars rather than a single massive seed black hole\,\citep{Chon2020MNRAS}. An alternative and increasingly favored pathway is the formation of IMBHs with masses of $\sim10^{3}$--$10^{4}\,M_\odot$ through runaway stellar collisions in dense star clusters\,\citep[e.g.,][]{Zwart2004Natur,Rantala2024MNRAS,Fujii2024Sci,Vergara2025arXiv,Mestichelli2026arXiv}. These considerations suggest that forming $\sim10^{3}\,M_\odot$ IMBH seeds may be more plausible than directly forming $\sim10^{5}\,M_\odot$ heavy seeds.

Due to the substantial computational cost of direct $N$-body simulations, our current calculations are restricted to IMBH masses up to $3\times10^4\,M_\odot$, and we are unable to directly follow the dTDE evolution at larger black hole masses or on longer timescales. As a first step, we have performed simulations for only three discrete IMBH masses, from which we measure the dTDE rates: $\dot{M}_{\rm dTDE} = 7.9\times10^{-4}\,M_\odot\,\mathrm{yr}^{-1}$ for $M_\bullet=3\times10^3\,M_\odot$, $7.8\times10^{-4}\,M_\odot\,\mathrm{yr}^{-1}$ for $M_\bullet=1\times10^4\,M_\odot$, and $2.6\times10^{-3}\,M_\odot\,\mathrm{yr}^{-1}$ for $M_\bullet=3\times10^4\,M_\odot$ (see Figure\,\ref{fig:BH_growth}). For IMBH masses within this range, we employ linear interpolation in the $\log M_\bullet$--$\log \dot{M}_{\rm dTDE}$ plane based on these three data points. For black hole masses beyond $3\times10^4\,M_\odot$, we adopt a constant dTDE rate of $2.6\times10^{-4}\,M_\odot\,\mathrm{yr}^{-1}$ as a conservative extrapolation. This assumption is motivated by two physical considerations. First, the dTDE rate is not expected to increase indefinitely with black hole mass. As the black hole becomes more massive, the timescale for capturing stars through star--disk interactions becomes longer, preventing a continuous increase of the dTDE rate with black hole mass\,\citep{wangyh2024ApJ}. Second, the Eddington accretion rate of the surrounding gas disk increases with black hole mass, eventually making gas accretion the dominant growth channel and reducing the relative contribution of dTDEs\,\citep[see][]{Kennedy2016MNRAS}.

In Figure~\ref{fig:BH_growth}, we present illustrative growth histories for seed IMBHs with an initial mass of $3\times10^3\,M_\odot$ formed at redshifts $z=25$, $20$, $15$, and $10$. 
The dTDE contribution is calculated using the mass-dependent dTDE rate obtained from our simulations, while the dashed curves show the corresponding growth under Eddington-limited gas accretion alone. 
The triangles indicate the masses of observed high-redshift SMBHs at their corresponding redshifts\,\citep[see][and references therein]{Taylor2025ApJ}. 
During the dTDE-assisted phase, we assume that the nuclear environment continuously supplies both stars and gas such that the compact cluster and accretion disk persist until the black hole reaches $\sim10^5\,M_\odot$. 
This assumption specifies the environmental requirement of the scenario. 
Beyond this stage, the contribution of dTDEs becomes subdominant and the subsequent growth is governed primarily by gas accretion.

As shown in Figure~\ref{fig:BH_growth}, Eddington-limited accretion alone results in relatively slow growth from a $3\times10^3\,M_\odot$ seed. 
In contrast, adding the simulated dTDE contribution accelerates the early growth phase and can raise the IMBH mass to $\gtrsim10^5\,M_\odot$ on a short timescale\,($\sim$ 30 Myr). 
Once this mass scale is reached, gas accretion becomes the dominant growth channel. 
Together, these curves show that dTDEs can act as an early accelerator of IMBH growth under favorable gas-rich cluster conditions, after which ordinary gas accretion controls the subsequent SMBH growth.

We note that this $\sim30\,\mathrm{Myr}$ growth timescale is not negligible compared with the typical timescales of star formation and feedback in high-redshift galaxies. 
It implicitly assumes that the natal accretion disk can be maintained over this entire period. 
Whether such long-lived, gas-rich nuclear disks are common in high-redshift galaxies, and under what conditions they can be sustained, remain open questions that require dedicated large-scale cosmological simulations with sufficient resolution to follow the interplay between gas inflow, star formation, and BH accretion. 
Such simulations are essential to robustly assess the overall contribution of the dTDE channel to the early growth of SMBHs.

\section{Conclusions and Discussions} \label{sec:discussion}

In this work, we investigate dTDEs as a stellar feeding channel for IMBHs embedded in gas-rich dense nuclear star clusters. 
By performing high-precision direct $N$-body simulations with analytic prescriptions for star--disk interactions, we follow the coupled evolution of the stellar cusp, relaxation-driven replenishment, disk capture, inward migration, stellar mergers, and eventual accretion onto the central IMBH. 
We find that this process can sustain stellar disruption rates of $\sim\,(1-3)\times 10^{-3}\,M_\odot\,\mathrm{yr}^{-1}$, significantly exceeding the Eddington accretion rate for black holes with $M_\bullet<10^5\,M_\odot$. 
The main conclusion is therefore that disk-induced stellar disruptions can provide an efficient and dynamically sustained stellar mass supply to IMBHs in compact gas-rich stellar systems. 
As a conditional application, this feeding channel can help grow $\sim 10^3\,M_\odot$ IMBHs toward $\gtrsim 10^5\,M_\odot$ if the required cluster and disk conditions are maintained.

\begin{figure}[htbp]
\centering
\includegraphics[scale=0.42]{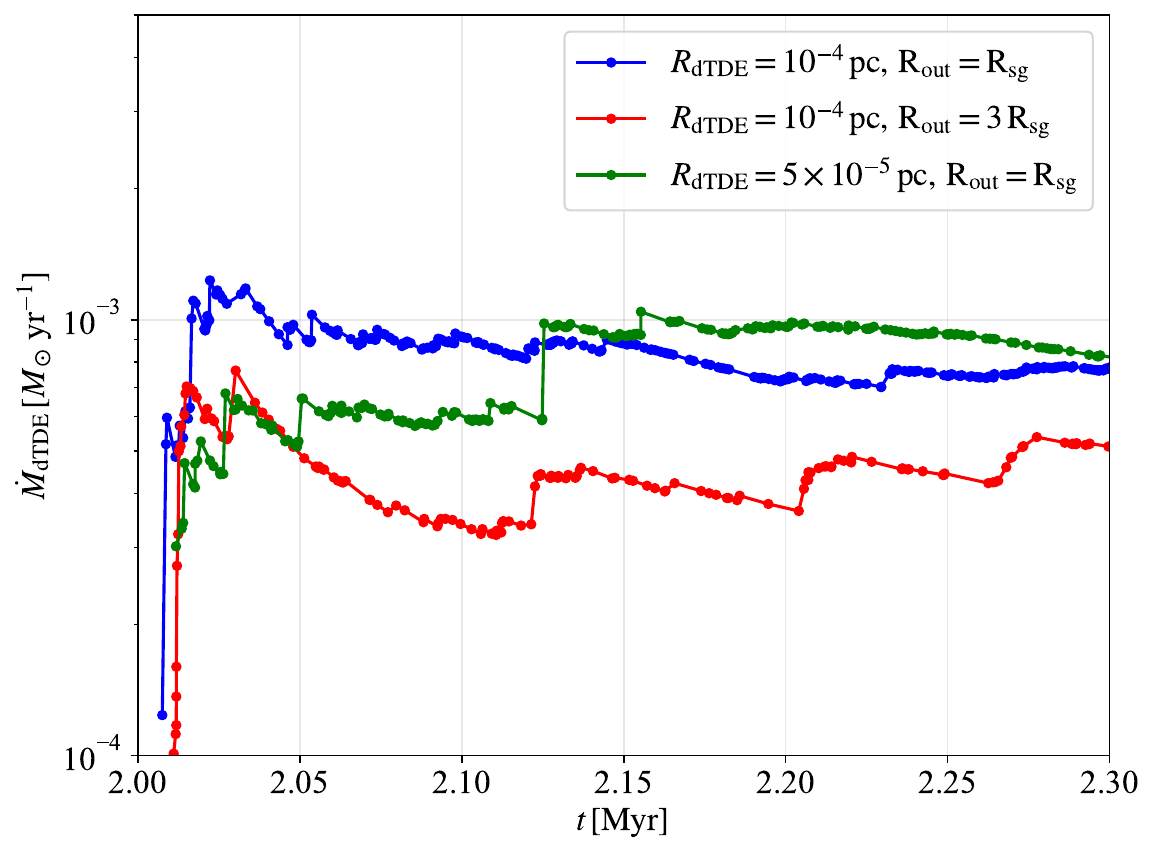}
\caption{Tests of the outer disk boundary and inner accretion radius for a black hole of mass $M_\bullet = 10^4\,M_\odot$. The blue line shows the fiducial case with $R_{\rm out} = R_{\rm sg}$ and $R_{\rm dTDE} = 10^{-4}\,\mathrm{pc}$. The red line uses a larger outer boundary $R_{\rm out} = 3\,R_{\rm sg}$ while keeping $R_{\rm dTDE} = 10^{-4}\,\mathrm{pc}$. The green line uses a smaller inner accretion radius $R_{\rm dTDE} = 5\times10^{-5}\,\mathrm{pc}$ while keeping $R_{\rm out} = R_{\rm sg}$.}
\label{fig:rate_test}
\end{figure}

However, several caveats and limitations of this work warrant discussion. 
These limitations are particularly important for interpreting the high-redshift application: our simulations test the local dynamical feeding mechanism, while the formation and lifetime of the required gas-rich nuclear environment must be supplied by galaxy-scale gas inflow, star formation, and feedback. 
A deeper assessment of how disk-induced TDEs affect seed black hole growth in a cosmological setting will therefore be addressed using cosmological simulations in a companion paper\,(Wang Z. et al. in preparation), which shows that dTDEs can boost IMBH growth to $\sim 10^5\,M_\odot$ within several tens of millions of years, increasing the abundance of massive black holes at $z\sim9$--$10$ and partially alleviating the tension with JWST-discovered overmassive candidates.

\subsection{Accretion Disk Model}
The outer radius of the accretion disk, $R_{\rm out}$, is a key parameter that determines the reservoir of stars available for capture. Theoretically, the disk structure is expected to consist of an inner, gravitationally stable standard accretion disk and an outer, marginally unstable star-forming disk\,\citep[][]{Sirko2003,TQM2005ApJ,Dittmann2020MNRAS}. As gas flows from galactic scales toward the central SMBH, the increasing density can drive the disk toward gravitational instability and fragmentation, triggering star formation. Feedback from newly formed stars\,(e.g., radiation and supernova explosions) can heat the gas and regulate further fragmentation, maintaining the disk in a marginally stable state with $Q\sim1$. The remaining gas continues to flow inward, feeding the inner accretion disk. However, the outer extent of this star-forming disk remains uncertain. Observations of nearby AGNs have not yet provided clear evidence for such an extended component, while recent JWST observations of ``little red dots'' with characteristic V-shaped spectra have been suggested as possible candidates for systems containing such compact star-forming disks in the early Universe\,\citep[][]{Zhang2026NatAs,Chen2026arXiv}.

In our fiducial model, we adopt $R_{\rm out}=R_{\rm sg}$, corresponding to the self-gravity radius of the disk. 
As shown in Figure~\ref{fig:rate_test}, extending the outer boundary to $R_{\rm out}=3R_{\rm sg}$\,(red curve) does not significantly alter the rate of dTDEs compared with the fiducial case\,(blue curve), but instead leads to a slight decrease from $\sim8\times10^{-4}$ to $\sim6\times10^{-4}\,M_\odot\,\mathrm{yr}^{-1}$. 
Although a larger $R_{\rm out}$ provides a larger population of stars that can interact with the disk, it also increases the frequency of two-body and few-body encounters, as well as stellar collisions within the disk. 
These additional dynamical interactions can perturb stellar orbits, increase their eccentricities and inclinations, and consequently reduce the efficiency of disk-driven inward migration. 
As a result, the overall dTDE rate remains only weakly dependent on the disk outer radius, suggesting that our conclusions are robust against reasonable variations in $R_{\rm out}$.

We further note that the structure of AGN accretion disks remains highly uncertain, with theoretical models spanning a wide range, from the canonical geometrically thin, thermal-pressure-dominated $\alpha$-disk to geometrically thicker configurations supported by radiation pressure, magnetic fields, or turbulence. Such geometrically thick disks are expected when these additional pressure components become dynamically important, for instance in radiation-pressure-dominated or strongly magnetized accretion flows. Variations in disk geometry can profoundly influence star--disk interactions. In particular, thick disks may suppress both the capture and subsequent inward migration of embedded stars, thereby significantly reducing the dTDE rate\,\citep{Tomar2026PhRvD}.

\subsection{Choice of inner accretion radius.}
Due to computational limitations, we replace the tidal disruption radius $R_{\rm t}\,(\sim 10^{-6} \rm pc)$ with a larger accretion radius $R_{\rm dTDE}$. To assess the robustness of this treatment, we compare our fiducial model\,($R_{\rm dTDE} = 10^{-4}\,\mathrm{pc}$) with a run adopting a smaller value, $R_{\rm dTDE} = 5\times10^{-5}\,\mathrm{pc}$\,(green curve), in Figure~\ref{fig:rate_test}. The resulting disruption rates are nearly identical, indicating that our results are insensitive to the specific choice of $R_{\rm dTDE}$. This insensitivity is physically expected: once stars are captured by the accretion disk, they settle onto nearly circular orbits and rapidly migrate inward. After crossing $R_{\rm dTDE}$, they continue migrating toward the central IMBH and typically reach the tidal disruption radius $R_{\rm t}$ within $\sim10^4$~yr. Therefore, as long as $R_{\rm dTDE}$ is sufficiently small, the resulting stellar disruption rate converges and is insensitive to the exact choice of $R_{\rm dTDE}$.

\subsection{Wet EMRI Contribution}
Our simulations span less than $3\,$Myr, which is too short for a significant stellar-mass black hole\,(sBH) population to form through stellar evolution. We therefore neglect the contribution of EMRIs to IMBH growth in this work. Even if sBHs were present, they would interact with the accretion disk and form wet EMRIs\,\citep[see][]{PanZhen2021wet,Wang2023MNRAS}, but their contribution is expected to be subdominant to the dTDE channel for three reasons. First, for our adopted Kroupa IMF, the number of sBHs is roughly $0.001$ times that of stars. Because disk capture and inward migration rates scale with mass, wet EMRIs contribute only $\sim 10\%$ of the mass growth achieved through dTDEs. Second, a star cluster with mass $2\times10^5\,M_\odot$ has an escape velocity of about $50$~km\,s$^{-1}$, whereas typical sBH natal kicks range from a few km\,s$^{-1}$ to $\gtrsim 100$~km\,s$^{-1}$ \citep[e.g.,][]{Nagarajan2025PASP}. Many sBHs are therefore ejected before they can interact with the disk. Third, for $M_\bullet\sim10^4\,M_\odot$, the gap-opening criterion (equation~\ref{eq:gap}) is satisfied for $m_\star\gtrsim3\,M_\odot$ in the inner disk. Consequently, sBHs with masses $\sim10\,M_\odot$ open a gap and migrate in the slower Type II regime, further suppressing their mass supply. A dedicated study of wet-EMRI formation and its gravitational-wave signatures at high redshift is deferred to future work.

\subsection{Stellar Collisions}
We treat physical collisions and mergers of stars using the standard ``sticky-sphere'' approximation implemented in the {\tt SDAR} module of {\tt PETAR}~\citep{WangLong2020MNRAS,WangLong2020MNRAS493}: two stars are merged when their pericenter distance falls below the sum of their radii, with the merger product conserving total mass and linear momentum. This treatment is admittedly simplified, since the outcome of a collision depends on both the impact speed and the impact parameter\,\citep{Lai1993ApJ,Freitag2005MNRAS,Rose2026ApJ}. In our simulations, the stellar collision velocities at contact are $\sim 500$--$1000~\mathrm{km\,s^{-1}}$, well above the $\sim 50~\mathrm{km\,s^{-1}}$ regime considered by \citet{Lai1993ApJ} and \citet{Freitag2005MNRAS}. At such high impact velocities, collisions may result in significant mass loss or may not result in a merger. These cases, however, do not significantly affect the dTDE rate for the following reasons. Recent SPH hydrodynamical simulations show that, for impact speeds below $\sim 10^{3}~\mathrm{km\,s^{-1}}$, the fractional mass loss in merging collisions is only $\sim10\%$--$20\%$\,\citep{Rose2026ApJ}. Therefore, for merging collisions, the mass-conserving sticky-sphere approximation is expected to differ only modestly from the hydrodynamical outcome. For collisions that do not result in a merger, the two stars continue to evolve independently, and each remains susceptible to tidal disruption by the IMBH. Consequently, the simplified treatment of mergers does not reduce the IMBH mass growth rate. Moreover, approximately $60\%$--$80\%$ of all stars tidally disrupted by the IMBH have not undergone any prior merger before disruption\,(see Figure\,\ref{fig:merger_statistics}). The dTDE rate is therefore insensitive to the details of the collision treatment.

\subsection{Other Limitations and future work.}
Despite the insights provided by our simulations, several other limitations should be acknowledged. 

Firstly, the conversion efficiency of disrupted stellar mass into black hole growth remains uncertain. Radiative hydrodynamic simulations of classical TDEs suggest accreted fractions of $\sim 25\%$--$43\%$ for a $10^6\,M_\odot$ SMBH disrupting stars of $1$--$300\,M_\odot$\,\citep{Bu2023MNRAS,Sheng2026ApJ}. In our disk-induced TDE scenario, however, stars are delivered on nearly circular, coplanar orbits and migrate inward through the disk prior to disruption. This may lead to more efficient debris circularization and thus a potentially higher accreted fraction than that achieved in the highly eccentric, parabolic orbits typical of classical TDEs. Nevertheless, our current assumption of full accretion should be regarded as an optimistic upper limit. 

Secondly, the limited duration of our simulations raises the question of whether the seemingly steady dTDE rate measured over $\sim 0.3-0.5$~Myr can be sustained over 30~Myr required to grow an IMBH to $\sim 10^5\,M_\odot$. This would require both a continuous external gas supply to maintain the disk and a sufficiently large stellar reservoir. We therefore assume a sustained large-scale gas injection as a prerequisite. Under this assumption, as the injected gas flows toward the central IMBH and its density increases, it undergoes gravitational collapse in a star-forming disk, leading to star formation\,\citep[e.g.,][]{TQM2005ApJ}. This large-scale gas injection provides a source for the persistence of the accretion disk and the continuous replenishment of stars in the stellar cluster.

Thirdly, our analytic treatment of star--disk interactions captures the essential dynamical processes, but we do not consider stellar feedback on the accretion disk, including disk crossing by cluster stars, and the migration, accretion, radiation, stellar winds, and supernovae of stars embedded in the disk. Such heating could affect the longevity and global morphology of the disk, and hence the migration rate of stars within it. \citet{Tagawa2020ApJ} discussed the effects of various heating processes on the accretion disk and found that the dominant mechanisms fall into three categories: (i) damping of the orbital velocities of stellar-mass objects as they cross the disk; (ii) migration torques for stellar-mass objects embedded in the disk; and (iii) radiation feedback from accretion onto sBHs. In our \(N\)-body simulations, about 300 stars have pericenters that cross the disk at a typical time-slice. Using equation~\eqref{eq:GDF}, we estimate the energy deposited into the disk by these stars through dynamical friction over one orbital period and find a total heating rate of \(\sim 5\times10^{38}\)~erg\,s\(^{-1} \simeq 4\times10^{-4}\,L_{\rm Edd}\). For embedded stars, inward migration from \(R_{\rm sg}\) to the tidal disruption radius \(R_{\rm t} = 10^{-6}\)~pc releases
\begin{equation}
\dot{E}_{\rm mig} \simeq \dot{M}_{\rm dTDE}\,\frac{G M_\bullet}{2}\left(\frac{1}{R_{\rm t}}-\frac{1}{R_{\rm sg}}\right)  \simeq 0.02\,L_{\rm Edd},
\end{equation}
where \(L_{\rm Edd} \simeq 1.3\times10^{42}\)~erg\,s\(^{-1}\) is the Eddington luminosity of the \(10^{4}\,M_\odot\) IMBH. Both channels are therefore far below the disk's radiative cooling luminosity and cannot affect its global energy budget or longevity. However, if a large population of sBHs is present in the disk, their accretion luminosity can significantly heat the disk, particularly at larger radii\,\citep[see][]{Tagawa2020ApJ}. Indeed, the stellar evolution, accretion, and feedback of stars embedded in AGN disks have recently emerged as a rapidly growing and active area of research, with potential observational consequences \citep[e.g.,][]{Huang2023MNRAS,CYX2023MNRAS,CYX2024ApJ,Xing2026MNRAS,Shi2026arXiv,Xu2026ApJ}. A more self-consistent treatment coupling these effects with radiation hydrodynamical and \(N\)-body simulations will be necessary to quantitatively assess their impact in future work.

\begin{acknowledgments}
We thank the referee for constructive comments and suggestions that helped improve this paper.
M.W. thanks Renyue Cen and Xiangli Lei for helpful discussions. Y.M. thanks Rongfeng Shen for discussions on the accretion fraction of tidally disrupted stars. Y.M, M.W. and Z.W thank Michele Mapelli for discussion and hosting their visit in Heidelberg.
M.W. and Q.W. gratefully acknowledge support from the National Natural Science Foundation of China\,(grants 123B2041 and 12233007), the National Key Research and Development Program of China\,(No. 2023YFC2206702) and the National SKA Program of China\,(2022SKA0120101). Y.M. is supported by the university start-up fund provided by Huazhong University of Science and Technology. The authors acknowledge Beijing PARATERA Technology Co., LTD for providing high-performance and AI computing resources for contributing to the research results.reported within this paper.
\end{acknowledgments}
\bibliography{ref}{}

@ARTICLE{Oshino2011PASJ,
       author = {{Oshino}, Shoichi and {Funato}, Yoko and {Makino}, Junichiro},
        title = "{Particle-Particle Particle-Tree: A Direct-Tree Hybrid Scheme for Collisional N-Body Simulations}",
      journal = {\pasj},
         year = 2011,
        month = aug,
       volume = {63},
        pages = {881},
          doi = {10.1093/pasj/63.4.881},
archivePrefix = {arXiv},
       eprint = {1101.5504},
 primaryClass = {astro-ph.EP},
       adsurl = {https://ui.adsabs.harvard.edu/abs/2011PASJ...63..881O}
}

@ARTICLE{Nagarajan2025PASP,
       author = {{Nagarajan}, Pranav and {El-Badry}, Kareem},
        title = "{Mixed Origins: Strong Natal Kicks for Some Black Holes and None for Others}",
      journal = {\pasp},
         year = 2025,
        month = mar,
       volume = {137},
       number = {3},
          eid = {034203},
        pages = {034203},
          doi = {10.1088/1538-3873/adb6d6},
archivePrefix = {arXiv},
       eprint = {2411.16847},
 primaryClass = {astro-ph.GA},
       adsurl = {https://ui.adsabs.harvard.edu/abs/2025PASP..137c4203N}
}

@ARTICLE{Freitag2005MNRAS,
       author = {{Freitag}, Marc and {Benz}, Willy},
        title = "{A comprehensive set of simulations of high-velocity collisions between main-sequence stars}",
      journal = {\mnras},
         year = 2005,
        month = apr,
       volume = {358},
       number = {4},
        pages = {1133-1158},
          doi = {10.1111/j.1365-2966.2005.08770.x},
archivePrefix = {arXiv},
       eprint = {astro-ph/0403621},
 primaryClass = {astro-ph},
       adsurl = {https://ui.adsabs.harvard.edu/abs/2005MNRAS.358.1133F}
}

@ARTICLE{Lai1993ApJ,
       author = {{Lai}, Dong and {Rasio}, Frederic A. and {Shapiro}, Stuart L.},
        title = "{Collisions and Close Encounters between Massive Main-Sequence Stars}",
      journal = {\apj},
         year = 1993,
        month = aug,
       volume = {412},
        pages = {593},
          doi = {10.1086/172946},
       adsurl = {https://ui.adsabs.harvard.edu/abs/1993ApJ...412..593L}
}

@ARTICLE{Rose2026ApJ,
       author = {{Rose}, Sanaea C. and {Lombardi}, Jr., James C. and {Gonz{\'a}lez Prieto}, Elena and {K{\i}ro{\u{g}}lu}, Fulya and {Rasio}, Frederic A.},
        title = "{Modeling Stellar Collisions in Galactic Nuclei Using Hydrodynamic Simulations and Machine Learning}",
      journal = {\apj},
         year = 2026,
        month = apr,
       volume = {1000},
       number = {2},
          eid = {162},
        pages = {162},
          doi = {10.3847/1538-4357/ae459b},
archivePrefix = {arXiv},
       eprint = {2511.01811},
 primaryClass = {astro-ph.GA},
       adsurl = {https://ui.adsabs.harvard.edu/abs/2026ApJ..1000..162R}
}

@ARTICLE{Stone2017MNRAS,
       author = {{Stone}, Nicholas C. and {K{\"u}pper}, Andreas H.~W. and {Ostriker}, Jeremiah P.},
        title = "{Formation of massive black holes in galactic nuclei: runaway tidal encounters}",
      journal = {\mnras},
         year = 2017,
        month = jun,
       volume = {467},
       number = {4},
        pages = {4180-4199},
          doi = {10.1093/mnras/stx097},
archivePrefix = {arXiv},
       eprint = {1606.01909},
 primaryClass = {astro-ph.GA},
       adsurl = {https://ui.adsabs.harvard.edu/abs/2017MNRAS.467.4180S}
}

@ARTICLE{Rose2022ApJ,
       author = {{Rose}, Sanaea C. and {Naoz}, Smadar and {Sari}, Re'em and {Linial}, Itai},
        title = "{The Formation of Intermediate-mass Black Holes in Galactic Nuclei}",
      journal = {\apjl},
         year = 2022,
        month = apr,
       volume = {929},
       number = {2},
          eid = {L22},
        pages = {L22},
          doi = {10.3847/2041-8213/ac6426},
archivePrefix = {arXiv},
       eprint = {2201.00022},
 primaryClass = {astro-ph.GA},
       adsurl = {https://ui.adsabs.harvard.edu/abs/2022ApJ...929L..22R}
}

@ARTICLE{Gonzalez2025ApJ,
       author = {{Gonz{\'a}lez Prieto}, Elena and {Rodriguez}, Carl L. and {Cabrera}, Tom{\'a}s},
        title = "{Growing the Intermediate-mass Black Hole in Omega Centauri}",
      journal = {\apjl},
         year = 2025,
        month = sep,
       volume = {990},
       number = {2},
          eid = {L69},
        pages = {L69},
          doi = {10.3847/2041-8213/adfd4a},
archivePrefix = {arXiv},
       eprint = {2507.06316},
 primaryClass = {astro-ph.HE},
       adsurl = {https://ui.adsabs.harvard.edu/abs/2025ApJ...990L..69G}
}

@ARTICLE{Huang2023MNRAS,
       author = {{Huang}, Jiamu and {Lin}, Douglas N.~C. and {Shields}, Gregory},
        title = "{Metal enrichment due to embedded stars in AGN discs}",
      journal = {\mnras},
         year = 2023,
        month = nov,
       volume = {525},
       number = {4},
        pages = {5702-5718},
          doi = {10.1093/mnras/stad2642},
archivePrefix = {arXiv},
       eprint = {2308.15761},
 primaryClass = {astro-ph.GA},
       adsurl = {https://ui.adsabs.harvard.edu/abs/2023MNRAS.525.5702H}
}

@ARTICLE{Shi2026arXiv,
       author = {{Shi}, Yanlong and {Huang}, Xiaoshan and {Lin}, Douglas N.~C. and {Murray}, Norman},
        title = "{Stellar mergers and chemical element mixing: implications for the metamorphic stellar evolution in AGN disks}",
      journal = {arXiv e-prints},
         year = 2026,
        month = jul,
          eid = {arXiv:2608.00242},
        pages = {arXiv:2608.00242},
          doi = {10.48550/arXiv.2608.00242},
archivePrefix = {arXiv},
       eprint = {2608.00242},
 primaryClass = {astro-ph.GA},
       adsurl = {https://ui.adsabs.harvard.edu/abs/2026arXiv260800242S}
}

@ARTICLE{Xu2026ApJ,
       author = {{Xu}, Zheng-Hao and {Chen}, Yi-Xian and {Lin}, Douglas N.~C.},
        title = "{Stellar Evolution with Radiative Feedback in AGN Disks}",
      journal = {\apj},
         year = 2026,
        month = feb,
       volume = {997},
       number = {2},
          eid = {206},
        pages = {206},
          doi = {10.3847/1538-4357/ae2271},
archivePrefix = {arXiv},
       eprint = {2511.03904},
 primaryClass = {astro-ph.GA},
       adsurl = {https://ui.adsabs.harvard.edu/abs/2026ApJ...997..206X}
}

@ARTICLE{CYX2024ApJ,
       author = {{Chen}, Yi-Xian and {Jiang}, Yan-Fei and {Goodman}, Jeremy and {Lin}, Douglas N.~C.},
        title = "{Radiation Hydrodynamic Simulations of Massive Stars in Gas-rich Environments: Accretion of AGN Stars Suppressed by Thermal Feedback}",
      journal = {\apj},
         year = 2024,
        month = oct,
       volume = {974},
       number = {1},
          eid = {106},
        pages = {106},
          doi = {10.3847/1538-4357/ad6dd4},
archivePrefix = {arXiv},
       eprint = {2408.12017},
 primaryClass = {astro-ph.HE},
       adsurl = {https://ui.adsabs.harvard.edu/abs/2024ApJ...974..106C}
}

@ARTICLE{CYX2023MNRAS,
       author = {{Chen}, Yi-Xian and {Lin}, Douglas N.~C.},
        title = "{Chaotic gas accretion by black holes embedded in AGN discs as cause of low-spin signatures in gravitational wave events}",
      journal = {\mnras},
         year = 2023,
        month = jun,
       volume = {522},
       number = {1},
        pages = {319-329},
          doi = {10.1093/mnras/stad992},
archivePrefix = {arXiv},
       eprint = {2303.17097},
 primaryClass = {astro-ph.HE},
       adsurl = {https://ui.adsabs.harvard.edu/abs/2023MNRAS.522..319C}
}

@ARTICLE{Xing2026MNRAS,
       author = {{Xing}, Jing-Tong and {Liu}, Tong and {She}, Jiao-Zhen},
        title = "{Reincarnations of massive stars in active galactic nucleus discs}",
      journal = {\mnras},
         year = 2026,
        month = sep,
       volume = {551},
       number = {2},
          eid = {stag1459},
        pages = {stag1459},
          doi = {10.1093/mnras/stag1459},
archivePrefix = {arXiv},
       eprint = {2607.29075},
 primaryClass = {astro-ph.GA},
       adsurl = {https://ui.adsabs.harvard.edu/abs/2026MNRAS.551g1459X}
}

@ARTICLE{Hurley2000MNRAS,
       author = {{Hurley}, Jarrod R. and {Pols}, Onno R. and {Tout}, Christopher A.},
        title = "{Comprehensive analytic formulae for stellar evolution as a function of mass and metallicity}",
      journal = {\mnras},
         year = 2000,
        month = jul,
       volume = {315},
       number = {3},
        pages = {543-569},
          doi = {10.1046/j.1365-8711.2000.03426.x},
archivePrefix = {arXiv},
       eprint = {astro-ph/0001295},
 primaryClass = {astro-ph},
       adsurl = {https://ui.adsabs.harvard.edu/abs/2000MNRAS.315..543H}
}

@ARTICLE{Hurley2002MNRAS,
       author = {{Hurley}, Jarrod R. and {Tout}, Christopher A. and {Pols}, Onno R.},
        title = "{Evolution of binary stars and the effect of tides on binary populations}",
      journal = {\mnras},
         year = 2002,
        month = feb,
       volume = {329},
       number = {4},
        pages = {897-928},
          doi = {10.1046/j.1365-8711.2002.05038.x},
archivePrefix = {arXiv},
       eprint = {astro-ph/0201220},
 primaryClass = {astro-ph},
       adsurl = {https://ui.adsabs.harvard.edu/abs/2002MNRAS.329..897H}
}

@ARTICLE{Neumayer2020A&ARv,
       author = {{Neumayer}, Nadine and {Seth}, Anil and {B{\"o}ker}, Torsten},
        title = "{Nuclear star clusters}",
      journal = {\aapr},
         year = 2020,
        month = jul,
       volume = {28},
       number = {1},
          eid = {4},
        pages = {4},
          doi = {10.1007/s00159-020-00125-0},
archivePrefix = {arXiv},
       eprint = {2001.03626},
 primaryClass = {astro-ph.GA},
       adsurl = {https://ui.adsabs.harvard.edu/abs/2020A&ARv..28....4N}
}

@ARTICLE{WZJ2025ApJ,
       author = {{Wang}, Zijian and {Ma}, Yiqiu and {Li}, Yuxuan and {Cai}, Zheng and {Wang}, Chanyan and {Wu}, Qingwen},
        title = "{The Role of Population III Star Tidal Disruption Events in Black Hole Growth at the Cosmic Dawn}",
      journal = {\apj},
         year = 2025,
        month = sep,
       volume = {990},
       number = {2},
          eid = {160},
        pages = {160},
          doi = {10.3847/1538-4357/adf435},
archivePrefix = {arXiv},
       eprint = {2504.18144},
 primaryClass = {astro-ph.HE},
       adsurl = {https://ui.adsabs.harvard.edu/abs/2025ApJ...990..160W}
}

@ARTICLE{Rees1988Natur,
       author = {{Rees}, Martin J.},
        title = "{Tidal disruption of stars by black holes of {}10$^{6}$-{}10$^{8}$ solar masses in nearby galaxies}",
      journal = {\nat},
         year = 1988,
        month = jun,
       volume = {333},
       number = {6173},
        pages = {523-528},
          doi = {10.1038/333523a0},
       adsurl = {https://ui.adsabs.harvard.edu/abs/1988Natur.333..523R}
}

@ARTICLE{Gezari2021ARA&A,
       author = {{Gezari}, Suvi},
        title = "{Tidal Disruption Events}",
      journal = {\araa},
         year = 2021,
        month = sep,
       volume = {59},
        pages = {21-58},
          doi = {10.1146/annurev-astro-111720-030029},
archivePrefix = {arXiv},
       eprint = {2104.14580},
 primaryClass = {astro-ph.HE},
       adsurl = {https://ui.adsabs.harvard.edu/abs/2021ARA&A..59...21G}
}

@ARTICLE{Kormendy2013ARA&A,
       author = {{Kormendy}, John and {Ho}, Luis C.},
        title = "{Coevolution (Or Not) of Supermassive Black Holes and Host Galaxies}",
      journal = {\araa},
         year = 2013,
        month = aug,
       volume = {51},
       number = {1},
        pages = {511-653},
          doi = {10.1146/annurev-astro-082708-101811},
archivePrefix = {arXiv},
       eprint = {1304.7762},
 primaryClass = {astro-ph.CO},
       adsurl = {https://ui.adsabs.harvard.edu/abs/2013ARA&A..51..511K}
}

@ARTICLE{Plummer1911MNRAS,
       author = {{Plummer}, H.~C.},
        title = "{On the problem of distribution in globular star clusters}",
      journal = {\mnras},
         year = 1911,
        month = mar,
       volume = {71},
        pages = {460-470},
          doi = {10.1093/mnras/71.5.460},
       adsurl = {https://ui.adsabs.harvard.edu/abs/1911MNRAS..71..460P}
}

@ARTICLE{Ostriker1999ApJ,
       author = {{Ostriker}, Eve C.},
        title = "{Dynamical Friction in a Gaseous Medium}",
      journal = {\apj},
         year = 1999,
        month = mar,
       volume = {513},
       number = {1},
        pages = {252-258},
          doi = {10.1086/306858},
archivePrefix = {arXiv},
       eprint = {astro-ph/9810324},
 primaryClass = {astro-ph},
       adsurl = {https://ui.adsabs.harvard.edu/abs/1999ApJ...513..252O}
}

@ARTICLE{Adachi1976PThPh,
       author = {{Adachi}, I. and {Hayashi}, C. and {Nakazawa}, K.},
        title = "{The gas drag effect on the elliptical motion of a solid body in the primordial solar nebula.}",
      journal = {Progress of Theoretical Physics},
         year = 1976,
        month = dec,
       volume = {56},
        pages = {1756-1771},
          doi = {10.1143/PTP.56.1756},
       adsurl = {https://ui.adsabs.harvard.edu/abs/1976PThPh..56.1756A}
}

@ARTICLE{Cresswell2008A&A,
       author = {{Cresswell}, P. and {Nelson}, R.~P.},
        title = "{Three-dimensional simulations of multiple protoplanets embedded in a protostellar disc}",
      journal = {\aap},
         year = 2008,
        month = may,
       volume = {482},
       number = {2},
        pages = {677-690},
          doi = {10.1051/0004-6361:20079178},
archivePrefix = {arXiv},
       eprint = {0811.4322},
 primaryClass = {astro-ph},
       adsurl = {https://ui.adsabs.harvard.edu/abs/2008A&A...482..677C}
}

@ARTICLE{Haiman2001ApJ,
       author = {{Haiman}, Zolt{\'a}n and {Loeb}, Abraham},
        title = "{What Is the Highest Plausible Redshift of Luminous Quasars?}",
      journal = {\apj},
         year = 2001,
        month = may,
       volume = {552},
       number = {2},
        pages = {459-463},
          doi = {10.1086/320586},
archivePrefix = {arXiv},
       eprint = {astro-ph/0011529},
 primaryClass = {astro-ph},
       adsurl = {https://ui.adsabs.harvard.edu/abs/2001ApJ...552..459H}
}

@ARTICLE{Madau2001ApJ,
       author = {{Madau}, Piero and {Rees}, Martin J.},
        title = "{Massive Black Holes as Population III Remnants}",
      journal = {\apjl},
         year = 2001,
        month = apr,
       volume = {551},
       number = {1},
        pages = {L27-L30},
          doi = {10.1086/319848},
archivePrefix = {arXiv},
       eprint = {astro-ph/0101223},
 primaryClass = {astro-ph},
       adsurl = {https://ui.adsabs.harvard.edu/abs/2001ApJ...551L..27M}
}

@ARTICLE{Oh2002ApJ,
       author = {{Oh}, S. Peng and {Haiman}, Zolt{\'a}n},
        title = "{Second-Generation Objects in the Universe: Radiative Cooling and Collapse of Halos with Virial Temperatures above {}10$^{4}$ K}",
      journal = {\apj},
         year = 2002,
        month = apr,
       volume = {569},
       number = {2},
        pages = {558-572},
          doi = {10.1086/339393},
archivePrefix = {arXiv},
       eprint = {astro-ph/0108071},
 primaryClass = {astro-ph},
       adsurl = {https://ui.adsabs.harvard.edu/abs/2002ApJ...569..558O}
}

@ARTICLE{Bromm2003ApJ,
       author = {{Bromm}, Volker and {Loeb}, Abraham},
        title = "{Formation of the First Supermassive Black Holes}",
      journal = {\apj},
         year = 2003,
        month = oct,
       volume = {596},
       number = {1},
        pages = {34-46},
          doi = {10.1086/377529},
archivePrefix = {arXiv},
       eprint = {astro-ph/0212400},
 primaryClass = {astro-ph},
       adsurl = {https://ui.adsabs.harvard.edu/abs/2003ApJ...596...34B}
}

@ARTICLE{Begelman2006MNRAS,
       author = {{Begelman}, Mitchell C. and {Volonteri}, Marta and {Rees}, Martin J.},
        title = "{Formation of supermassive black holes by direct collapse in pre-galactic haloes}",
      journal = {\mnras},
         year = 2006,
        month = jul,
       volume = {370},
       number = {1},
        pages = {289-298},
          doi = {10.1111/j.1365-2966.2006.10467.x},
archivePrefix = {arXiv},
       eprint = {astro-ph/0602363},
 primaryClass = {astro-ph},
       adsurl = {https://ui.adsabs.harvard.edu/abs/2006MNRAS.370..289B}
}

@ARTICLE{Zwart2004Natur,
       author = {{Portegies Zwart}, Simon F. and {Baumgardt}, Holger and {Hut}, Piet and {Makino}, Junichiro and {McMillan}, Stephen L.~W.},
        title = "{Formation of massive black holes through runaway collisions in dense young star clusters}",
      journal = {\nat},
         year = 2004,
        month = apr,
       volume = {428},
       number = {6984},
        pages = {724-726},
          doi = {10.1038/nature02448},
archivePrefix = {arXiv},
       eprint = {astro-ph/0402622},
 primaryClass = {astro-ph},
       adsurl = {https://ui.adsabs.harvard.edu/abs/2004Natur.428..724P}
}

@ARTICLE{Vergara2025arXiv,
       author = {{Vergara}, Marcelo C. and {Askar}, Abbas and {Kamlah}, Albrecht W.~H. and {Spurzem}, Rainer and {Flammini Dotti}, Francesco and {Schleicher}, Dominik R.~G. and {Arca Sedda}, Manuel and {Hypki}, Arkadiusz and {Giersz}, Mirek and {Hurley}, Jarrod and {Berczik}, Peter and {Escala}, Andres and {Hoyer}, Nils and {Neumayer}, Nadine and {Pang}, Xiaoying and {Tanikawa}, Ataru and {Cen}, Renyue and {Naab}, Thorsten},
        title = "{Rapid formation of a very massive star >50000 $M_\odot$ and subsequently an IMBH from runaway collisions. Direct N-body and Monte Carlo simulations of dense star clusters}",
      journal = {arXiv e-prints},
         year = 2025,
        month = may,
          eid = {arXiv:2505.07491},
        pages = {arXiv:2505.07491},
          doi = {10.48550/arXiv.2505.07491},
archivePrefix = {arXiv},
       eprint = {2505.07491},
 primaryClass = {astro-ph.GA},
       adsurl = {https://ui.adsabs.harvard.edu/abs/2025arXiv250507491V}
}

@ARTICLE{TQM2005ApJ,
       author = {{Thompson}, Todd A. and {Quataert}, Eliot and {Murray}, Norman},
        title = "{Radiation Pressure-supported Starburst Disks and Active Galactic Nucleus Fueling}",
      journal = {\apj},
         year = 2005,
        month = sep,
       volume = {630},
       number = {1},
        pages = {167-185},
          doi = {10.1086/431923},
archivePrefix = {arXiv},
       eprint = {astro-ph/0503027},
 primaryClass = {astro-ph},
       adsurl = {https://ui.adsabs.harvard.edu/abs/2005ApJ...630..167T}
}

@ARTICLE{Dittmann2020MNRAS,
       author = {{Dittmann}, Alexander J. and {Miller}, M. Coleman},
        title = "{Star formation in accretion discs and SMBH growth}",
      journal = {\mnras},
         year = 2020,
        month = apr,
       volume = {493},
       number = {3},
        pages = {3732-3743},
          doi = {10.1093/mnras/staa463},
archivePrefix = {arXiv},
       eprint = {1911.08685},
 primaryClass = {astro-ph.HE},
       adsurl = {https://ui.adsabs.harvard.edu/abs/2020MNRAS.493.3732D}
}

@ARTICLE{Zhang2026NatAs,
       author = {{Zhang}, Chenxuan and {Wu}, Qingwen and {Fan}, Xiao and {Ho}, Luis C. and {Wu}, Jiancheng and {Zhang}, Huanian and {Lyu}, Bing and {Cao}, Xinwu and {Wang}, Jianmin},
        title = "{The composite spectrum of little red dots from a standard inner disk and an unstable outer disk}",
      journal = {Nature Astronomy},
         year = 2026,
        month = feb,
          doi = {10.1038/s41550-026-02785-x},
archivePrefix = {arXiv},
       eprint = {2505.12719},
 primaryClass = {astro-ph.HE},
       adsurl = {https://ui.adsabs.harvard.edu/abs/2026NatAs.tmp...41Z}
}

@ARTICLE{Chen2026arXiv,
       author = {{Chen}, Yi-Xian and {Liu}, Hanpu and {Li}, Ruancun and {Wang}, Bingjie and {Ma}, Yilun and {Jiang}, Yan-Fei and {Greene}, Jenny E. and {Quataert}, Eliot and {Goodman}, Jeremy},
        title = "{Spectral Appearance of Self-gravitating AGN Disks Powered by Stellar Objects: Universal Effective Temperature in the Optical Continuum and Application to Little Red Dots}",
      journal = {arXiv e-prints},
         year = 2026,
        month = feb,
          eid = {arXiv:2602.06954},
        pages = {arXiv:2602.06954},
          doi = {10.48550/arXiv.2602.06954},
archivePrefix = {arXiv},
       eprint = {2602.06954},
 primaryClass = {astro-ph.HE},
       adsurl = {https://ui.adsabs.harvard.edu/abs/2026arXiv260206954C}
}

@ARTICLE{Latif2016ApJ,
       author = {{Latif}, M.~A. and {Omukai}, K. and {Habouzit}, M. and {Schleicher}, D.~R.~G. and {Volonteri}, M.},
        title = "{Impact of Dust Cooling on Direct-collapse Black Hole Formation}",
      journal = {\apj},
         year = 2016,
        month = may,
       volume = {823},
       number = {1},
          eid = {40},
        pages = {40},
          doi = {10.3847/0004-637X/823/1/40},
archivePrefix = {arXiv},
       eprint = {1509.07034},
 primaryClass = {astro-ph.GA},
       adsurl = {https://ui.adsabs.harvard.edu/abs/2016ApJ...823...40L}
}

@ARTICLE{Omukai2008ApJ,
       author = {{Omukai}, K. and {Schneider}, R. and {Haiman}, Z.},
        title = "{Can Supermassive Black Holes Form in Metal-enriched High-Redshift Protogalaxies?}",
      journal = {\apj},
         year = 2008,
        month = oct,
       volume = {686},
       number = {2},
        pages = {801-814},
          doi = {10.1086/591636},
archivePrefix = {arXiv},
       eprint = {0804.3141},
 primaryClass = {astro-ph},
       adsurl = {https://ui.adsabs.harvard.edu/abs/2008ApJ...686..801O}
}

@ARTICLE{Chon2020MNRAS,
       author = {{Chon}, Sunmyon and {Omukai}, Kazuyuki},
        title = "{Supermassive star formation via super competitive accretion in slightly metal-enriched clouds}",
      journal = {\mnras},
         year = 2020,
        month = may,
       volume = {494},
       number = {2},
        pages = {2851-2860},
          doi = {10.1093/mnras/staa863},
archivePrefix = {arXiv},
       eprint = {2001.06491},
 primaryClass = {astro-ph.GA},
       adsurl = {https://ui.adsabs.harvard.edu/abs/2020MNRAS.494.2851C}
}

@ARTICLE{Fujii2024Sci,
       author = {{Fujii}, Michiko S. and {Wang}, Long and {Tanikawa}, Ataru and {Hirai}, Yutaka and {Saitoh}, Takayuki R.},
        title = "{Simulations predict intermediate-mass black hole formation in globular clusters}",
      journal = {Science},
         year = 2024,
        month = jun,
       volume = {384},
       number = {6703},
        pages = {1488-1492},
          doi = {10.1126/science.adi4211},
archivePrefix = {arXiv},
       eprint = {2406.06772},
 primaryClass = {astro-ph.GA},
       adsurl = {https://ui.adsabs.harvard.edu/abs/2024Sci...384.1488F}
}

@ARTICLE{Rantala2024MNRAS,
       author = {{Rantala}, Antti and {Naab}, Thorsten and {Lah{\'e}n}, Natalia},
        title = "{FROST-CLUSTERS - I. Hierarchical star cluster assembly boosts intermediate-mass black hole formation}",
      journal = {\mnras},
         year = 2024,
        month = jul,
       volume = {531},
       number = {3},
        pages = {3770-3799},
          doi = {10.1093/mnras/stae1413},
archivePrefix = {arXiv},
       eprint = {2403.10602},
 primaryClass = {astro-ph.GA},
       adsurl = {https://ui.adsabs.harvard.edu/abs/2024MNRAS.531.3770R}
}

@ARTICLE{Alexander2025NewAR,
       author = {{Alexander}, D.~M. and {Hickox}, R.~C. and {Aird}, J. and {Combes}, F. and {Costa}, T. and {Habouzit}, M. and {Harrison}, C.~M. and {Leng}, R.~I. and {Morabito}, L.~K. and {Uckelman}, S.~L. and {Vickers}, P.},
        title = "{What drives the growth of black holes: A decade of progress}",
      journal = {\nar},
         year = 2025,
        month = dec,
       volume = {101},
          eid = {101733},
        pages = {101733},
          doi = {10.1016/j.newar.2025.101733},
archivePrefix = {arXiv},
       eprint = {2506.19166},
 primaryClass = {astro-ph.GA},
       adsurl = {https://ui.adsabs.harvard.edu/abs/2025NewAR.10101733A}
}

@ARTICLE{Furtak2024Natur,
       author = {{Furtak}, Lukas J. and {Labb{\'e}}, Ivo and {Zitrin}, Adi and {Greene}, Jenny E. and {Dayal}, Pratika and {Chemerynska}, Iryna and {Kokorev}, Vasily and {Miller}, Tim B. and {Goulding}, Andy D. and {de Graaff}, Anna and {Bezanson}, Rachel and {Brammer}, Gabriel B. and {Cutler}, Sam E. and {Leja}, Joel and {Pan}, Richard and {Price}, Sedona H. and {Wang}, Bingjie and {Weaver}, John R. and {Whitaker}, Katherine E. and {Atek}, Hakim and {Bogd{\'a}n}, {\'A}kos and {Charlot}, St{\'e}phane and {Curtis-Lake}, Emma and {van Dokkum}, Pieter and {Endsley}, Ryan and {Feldmann}, Robert and {Fudamoto}, Yoshinobu and {Fujimoto}, Seiji and {Glazebrook}, Karl and {Juneau}, St{\'e}phanie and {Marchesini}, Danilo and {Maseda}, Micheal V. and {Nelson}, Erica and {Oesch}, Pascal A. and {Plat}, Ad{\`e}le and {Setton}, David J. and {Stark}, Daniel P. and {Williams}, Christina C.},
        title = "{A high black-hole-to-host mass ratio in a lensed AGN in the early Universe}",
      journal = {\nat},
         year = 2024,
        month = apr,
       volume = {628},
       number = {8006},
        pages = {57-61},
          doi = {10.1038/s41586-024-07184-8},
archivePrefix = {arXiv},
       eprint = {2308.05735},
 primaryClass = {astro-ph.GA},
       adsurl = {https://ui.adsabs.harvard.edu/abs/2024Natur.628...57F}
}

@ARTICLE{Maiolino2024A&A,
       author = {{Maiolino}, Roberto and {Scholtz}, Jan and {Curtis-Lake}, Emma and {Carniani}, Stefano and {Baker}, William and {de Graaff}, Anna and {Tacchella}, Sandro and {{\"U}bler}, Hannah and {D'Eugenio}, Francesco and {Witstok}, Joris and {Curti}, Mirko and {Arribas}, Santiago and {Bunker}, Andrew J. and {Charlot}, St{\'e}phane and {Chevallard}, Jacopo and {Eisenstein}, Daniel J. and {Egami}, Eiichi and {Ji}, Zhiyuan and {Jones}, Gareth C. and {Lyu}, Jianwei and {Rawle}, Tim and {Robertson}, Brant and {Rujopakarn}, Wiphu and {Perna}, Michele and {Sun}, Fengwu and {Venturi}, Giacomo and {Williams}, Christina C. and {Willott}, Chris},
        title = "{JADES: The diverse population of infant black holes at 4 < z < 11: Merging, tiny, poor, but mighty}",
      journal = {\aap},
         year = 2024,
        month = nov,
       volume = {691},
          eid = {A145},
        pages = {A145},
          doi = {10.1051/0004-6361/202347640},
archivePrefix = {arXiv},
       eprint = {2308.01230},
 primaryClass = {astro-ph.GA},
       adsurl = {https://ui.adsabs.harvard.edu/abs/2024A&A...691A.145M}
}

@ARTICLE{Inayoshi2020ARA&A,
       author = {{Inayoshi}, Kohei and {Visbal}, Eli and {Haiman}, Zolt{\'a}n},
        title = "{The Assembly of the First Massive Black Holes}",
      journal = {\araa},
         year = 2020,
        month = aug,
       volume = {58},
        pages = {27-97},
          doi = {10.1146/annurev-astro-120419-014455},
archivePrefix = {arXiv},
       eprint = {1911.05791},
 primaryClass = {astro-ph.GA},
       adsurl = {https://ui.adsabs.harvard.edu/abs/2020ARA&A..58...27I}
}

@ARTICLE{Fan2023ARA&A,
       author = {{Fan}, Xiaohui and {Ba{\~n}ados}, Eduardo and {Simcoe}, Robert A.},
        title = "{Quasars and the Intergalactic Medium at Cosmic Dawn}",
      journal = {\araa},
         year = 2023,
        month = aug,
       volume = {61},
        pages = {373-426},
          doi = {10.1146/annurev-astro-052920-102455},
archivePrefix = {arXiv},
       eprint = {2212.06907},
 primaryClass = {astro-ph.GA},
       adsurl = {https://ui.adsabs.harvard.edu/abs/2023ARA&A..61..373F}
}

@ARTICLE{Taylor2025ApJ,
       author = {{Taylor}, Anthony J. and {Kokorev}, Vasily and {Kocevski}, Dale D. and {Akins}, Hollis B. and {Cullen}, Fergus and {Dickinson}, Mark and {Finkelstein}, Steven L. and {Arrabal Haro}, Pablo and {Bromm}, Volker and {Giavalisco}, Mauro and {Inayoshi}, Kohei and {Juneau}, St{\'e}phanie and {Leung}, Gene C.~K. and {P{\'e}rez-Gonz{\'a}lez}, Pablo G. and {Somerville}, Rachel S. and {Trump}, Jonathan R. and {Amor{\'\i}n}, Ricardo O. and {Barro}, Guillermo and {Burgarella}, Denis and {Brooks}, Madisyn and {Carnall}, Adam C. and {Casey}, Caitlin M. and {Cheng}, Yingjie and {Chisholm}, John and {Chworowsky}, Katherine and {Davis}, Kelcey and {Donnan}, Callum T. and {Dunlop}, James S. and {Ellis}, Richard S. and {Fern{\'a}ndez}, Vital and {Fujimoto}, Seiji and {Grogin}, Norman A. and {Gupta}, Ansh R. and {Hathi}, Nimish P. and {Jung}, Intae and {Hirschmann}, Michaela and {Kartaltepe}, Jeyhan S. and {Koekemoer}, Anton M. and {Larson}, Rebecca L. and {Leung}, Ho-Hin and {Llerena}, Mario and {Lucas}, Ray A. and {McLeod}, Derek J. and {McLure}, Ross and {Napolitano}, Lorenzo and {Papovich}, Casey and {Stanton}, Thomas M. and {Tripodi}, Roberta and {Wang}, Xin and {Wilkins}, Stephen M. and {Yung}, L.~Y. Aaron and {Zavala}, Jorge A.},
        title = "{CAPERS-LRD-z9: A Gas-enshrouded Little Red Dot Hosting a Broad-line Active Galactic Nucleus at z = 9.288}",
      journal = {\apjl},
         year = 2025,
        month = aug,
       volume = {989},
       number = {1},
          eid = {L7},
        pages = {L7},
          doi = {10.3847/2041-8213/ade789},
archivePrefix = {arXiv},
       eprint = {2505.04609},
 primaryClass = {astro-ph.GA},
       adsurl = {https://ui.adsabs.harvard.edu/abs/2025ApJ...989L...7T}
}

@ARTICLE{Bahcall1977ApJ,
       author = {{Bahcall}, J.~N. and {Wolf}, R.~A.},
        title = "{The star distribution around a massive black hole in a globular cluster. II. Unequal star masses.}",
      journal = {\apj},
         year = 1977,
        month = sep,
       volume = {216},
        pages = {883-907},
          doi = {10.1086/155534},
       adsurl = {https://ui.adsabs.harvard.edu/abs/1977ApJ...216..883B}
}

@ARTICLE{Bahcall1976ApJ,
       author = {{Bahcall}, J.~N. and {Wolf}, R.~A.},
        title = "{Star distribution around a massive black hole in a globular cluster.}",
      journal = {\apj},
         year = 1976,
        month = oct,
       volume = {209},
        pages = {214-232},
          doi = {10.1086/154711},
       adsurl = {https://ui.adsabs.harvard.edu/abs/1976ApJ...209..214B}
}

@ARTICLE{Crida2006Icar,
       author = {{Crida}, A. and {Morbidelli}, A. and {Masset}, F.},
        title = "{On the width and shape of gaps in protoplanetary disks}",
      journal = {\icarus},
         year = 2006,
        month = apr,
       volume = {181},
       number = {2},
        pages = {587-604},
          doi = {10.1016/j.icarus.2005.10.007},
archivePrefix = {arXiv},
       eprint = {astro-ph/0511082},
 primaryClass = {astro-ph},
       adsurl = {https://ui.adsabs.harvard.edu/abs/2006Icar..181..587C}
}

@ARTICLE{Lin1986,
       author = {{Lin}, D.~N.~C. and {Papaloizou}, John},
        title = "{On the Tidal Interaction between Protoplanets and the Protoplanetary Disk. III. Orbital Migration of Protoplanets}",
      journal = {\apj},
         year = 1986,
        month = oct,
       volume = {309},
        pages = {846},
          doi = {10.1086/164653},
       adsurl = {https://ui.adsabs.harvard.edu/abs/1986ApJ...309..846L}
}

@ARTICLE{Goldreich1980,
       author = {{Goldreich}, P. and {Tremaine}, S.},
        title = "{Disk-satellite interactions.}",
      journal = {\apj},
         year = 1980,
        month = oct,
       volume = {241},
        pages = {425-441},
          doi = {10.1086/158356},
       adsurl = {https://ui.adsabs.harvard.edu/abs/1980ApJ...241..425G}
}

@ARTICLE{wu2025ApJ,
       author = {{Wu}, Ziyong and {Cen}, Renyue and {Teyssier}, Romain},
        title = "{How Fast Could Supermassive Black Holes Grow at the Epoch of Reionization?}",
      journal = {\apjl},
         year = 2025,
        month = nov,
       volume = {993},
       number = {2},
          eid = {L48},
        pages = {L48},
          doi = {10.3847/2041-8213/ae14d4},
archivePrefix = {arXiv},
       eprint = {2510.16532},
 primaryClass = {astro-ph.GA},
       adsurl = {https://ui.adsabs.harvard.edu/abs/2025ApJ...993L..48W}
}

@ARTICLE{Wang2023MNRAS,
       author = {{Wang}, Mengye and {Ma}, Yiqiu and {Wu}, Qingwen},
        title = "{Accretion-modified stellar-mass black hole distribution and milli-Hz gravitational wave backgrounds from galaxy centre}",
      journal = {\mnras},
         year = 2023,
        month = apr,
       volume = {520},
       number = {3},
        pages = {4502-4516},
          doi = {10.1093/mnras/stad422},
archivePrefix = {arXiv},
       eprint = {2212.05724},
 primaryClass = {astro-ph.HE},
       adsurl = {https://ui.adsabs.harvard.edu/abs/2023MNRAS.520.4502W}
}

@article{Panzhen2021PhysRevD,
  title = {Formation rate of extreme mass ratio inspirals in active galactic nuclei},
  author = {Pan, Zhen and Yang, Huan},
  journal = {Phys. Rev. D},
  volume = {103},
  issue = {10},
  pages = {103018},
  numpages = {22},
  year = {2021},
  month = {May},
  publisher = {American Physical Society},
  doi = {10.1103/PhysRevD.103.103018},
  url = {https://link.aps.org/doi/10.1103/PhysRevD.103.103018}
}

@ARTICLE{PanZhen2021wet,
       author = {{Pan}, Zhen and {Lyu}, Zhenwei and {Yang}, Huan},
        title = "{Wet extreme mass ratio inspirals may be more common for spaceborne gravitational wave detection}",
      journal = {\prd},
         year = 2021,
        month = sep,
       volume = {104},
       number = {6},
          eid = {063007},
        pages = {063007},
          doi = {10.1103/PhysRevD.104.063007},
archivePrefix = {arXiv},
       eprint = {2104.01208},
 primaryClass = {astro-ph.HE},
       adsurl = {https://ui.adsabs.harvard.edu/abs/2021PhRvD.104f3007P}
}

@BOOK{Merritt2013book,
       author = {{Merritt}, David},
        title = "{Dynamics and Evolution of Galactic Nuclei}",
         year = 2013,
       adsurl = {https://ui.adsabs.harvard.edu/abs/2013degn.book.....M}
}

@ARTICLE{Sassano2023MNRAS,
       author = {{Sassano}, Federica and {Capelo}, Pedro R. and {Mayer}, Lucio and {Schneider}, Raffaella and {Valiante}, Rosa},
        title = "{Super-critical accretion of medium-weight seed black holes in gaseous proto-galactic nuclei}",
      journal = {\mnras},
         year = 2023,
        month = feb,
       volume = {519},
       number = {2},
        pages = {1837-1855},
          doi = {10.1093/mnras/stac3608},
archivePrefix = {arXiv},
       eprint = {2204.10330},
 primaryClass = {astro-ph.GA},
       adsurl = {https://ui.adsabs.harvard.edu/abs/2023MNRAS.519.1837S}
}

@ARTICLE{Shi2023MNRAS,
       author = {{Shi}, Yanlong and {Kremer}, Kyle and {Grudi{\'c}}, Michael Y. and {Gerling-Dunsmore}, Hannalore J. and {Hopkins}, Philip F.},
        title = "{Hyper-Eddington black hole growth in star-forming molecular clouds and galactic nuclei: can it happen?}",
      journal = {\mnras},
         year = 2023,
        month = jan,
       volume = {518},
       number = {3},
        pages = {3606-3621},
          doi = {10.1093/mnras/stac3245},
archivePrefix = {arXiv},
       eprint = {2208.05025},
 primaryClass = {astro-ph.GA},
       adsurl = {https://ui.adsabs.harvard.edu/abs/2023MNRAS.518.3606S}
}

@ARTICLE{Mestichelli2026arXiv,
       author = {{Mestichelli}, Benedetta and {Arca Sedda}, Manuel and {Volonteri}, Marta and {Mapelli}, Michela and {Torniamenti}, Stefano and {Lupi}, Alessandro and {Branchesi}, Marica and {Hirano}, Shingo and {Ishiyama}, Tomoaki and {Klessen}, Ralf S. and {Lipatova}, Veronika and {Liu}, Boyuan},
        title = "{Pebbles to Gems: Intermediate-mass black holes in the first star clusters}",
      journal = {arXiv e-prints},
         year = 2026,
        month = jul,
          eid = {arXiv:2607.03536},
        pages = {arXiv:2607.03536},
          doi = {10.48550/arXiv.2607.03536},
archivePrefix = {arXiv},
       eprint = {2607.03536},
 primaryClass = {astro-ph.GA},
       adsurl = {https://ui.adsabs.harvard.edu/abs/2026arXiv260703536M}
}

@ARTICLE{Brennan2025OJAp,
       author = {{O'Brennan}, Hannah and {Regan}, John A. and {Brennan}, John and {McCaffrey}, Joe and {Wise}, John H. and {Visbal}, Eli and {Trinca}, Alessandro and {Norman}, Michael L.},
        title = "{Predicting the number density of heavy seed massive black holes due to an intense Lyman-Werner field}",
      journal = {The Open Journal of Astrophysics},
         year = 2025,
        month = jul,
       volume = {8},
          eid = {88},
        pages = {88},
          doi = {10.33232/001c.141953},
archivePrefix = {arXiv},
       eprint = {2502.00574},
 primaryClass = {astro-ph.CO},
       adsurl = {https://ui.adsabs.harvard.edu/abs/2025OJAp....8E..88O}
}

@ARTICLE{Chon2026arXiv,
       author = {{Chon}, Sunmyon and {Hirano}, Shingo and {Ishiyama}, Tomoaki and {Chang}, Seok-Jun and {Springel}, Volker},
        title = "{Rapid emergence of overmassive black holes in the early Universe}",
      journal = {arXiv e-prints},
         year = 2026,
        month = jan,
          eid = {arXiv:2601.04955},
        pages = {arXiv:2601.04955},
          doi = {10.48550/arXiv.2601.04955},
archivePrefix = {arXiv},
       eprint = {2601.04955},
 primaryClass = {astro-ph.GA},
       adsurl = {https://ui.adsabs.harvard.edu/abs/2026arXiv260104955C}
}

@ARTICLE{Kimura2025ApJ,
       author = {{Kimura}, Kazutaka and {Inayoshi}, Kohei and {Omukai}, Kazuyuki},
        title = "{Massive Black Hole Seed Formation in Strong X-Ray Environments at High Redshift}",
      journal = {\apj},
         year = 2025,
        month = sep,
       volume = {990},
       number = {2},
          eid = {228},
        pages = {228},
          doi = {10.3847/1538-4357/adf2ad},
archivePrefix = {arXiv},
       eprint = {2504.10581},
 primaryClass = {astro-ph.GA},
       adsurl = {https://ui.adsabs.harvard.edu/abs/2025ApJ...990..228K}
}

@ARTICLE{Schleicher2022MNRAS,
       author = {{Schleicher}, D.~R.~G. and {Reinoso}, B. and {Latif}, M. and {Klessen}, R.~S. and {Vergara}, M.~Z.~C. and {Das}, A. and {Alister}, P. and {D{\'\i}az}, V.~B. and {Solar}, P.~A.},
        title = "{Origin of supermassive black holes in massive metal-poor protoclusters}",
      journal = {\mnras},
         year = 2022,
        month = jun,
       volume = {512},
       number = {4},
        pages = {6192-6200},
          doi = {10.1093/mnras/stac926},
archivePrefix = {arXiv},
       eprint = {2204.02361},
 primaryClass = {astro-ph.GA},
       adsurl = {https://ui.adsabs.harvard.edu/abs/2022MNRAS.512.6192S}
}

@ARTICLE{Cenci2025MNRAS,
       author = {{Cenci}, Elia and {Habouzit}, Melanie},
        title = "{Little Red Dots as direct-collapse black hole nurseries}",
      journal = {\mnras},
         year = 2025,
        month = sep,
       volume = {542},
       number = {3},
        pages = {2597-2609},
          doi = {10.1093/mnras/staf1362},
archivePrefix = {arXiv},
       eprint = {2508.14897},
 primaryClass = {astro-ph.GA},
       adsurl = {https://ui.adsabs.harvard.edu/abs/2025MNRAS.542.2597C}
}

@ARTICLE{Sheng2026ApJ,
       author = {{Sheng}, Yu-Heng and {Bu}, De-Fu and {Yang}, Xiao-Hong and {Chang}, Yi-Ren and {Chen}, Liang},
        title = "{Numerical Simulations of the Circularized Accretion Flow in Population III Star Tidal Disruption Events. I. The Accretion Flow and the Wind}",
      journal = {\apj},
         year = 2026,
        month = feb,
       volume = {998},
       number = {1},
          eid = {23},
        pages = {23},
          doi = {10.3847/1538-4357/ae3007},
archivePrefix = {arXiv},
       eprint = {2512.21500},
 primaryClass = {astro-ph.HE},
       adsurl = {https://ui.adsabs.harvard.edu/abs/2026ApJ...998...23S}
}

@ARTICLE{Bu2023MNRAS,
       author = {{Bu}, De-Fu and {Qiao}, Erlin and {Yang}, Xiao-Hong},
        title = "{Radiative hydrodynamical simulations of super-Eddington accretion flow in tidal disruption event: the accretion flow and wind}",
      journal = {\mnras},
         year = 2023,
        month = aug,
       volume = {523},
       number = {3},
        pages = {4136-4145},
          doi = {10.1093/mnras/stad1696},
archivePrefix = {arXiv},
       eprint = {2306.04313},
 primaryClass = {astro-ph.HE},
       adsurl = {https://ui.adsabs.harvard.edu/abs/2023MNRAS.523.4136B}
}

@ARTICLE{Kiyuna2026MNRAS,
       author = {{Kiyuna}, Masaki},
        title = "{Super-Eddington growth ceiling: analytic constraints on the rapid growth of light-seed black holes in massive clumps}",
      journal = {\mnras},
         year = 2026,
        month = jan,
       volume = {545},
       number = {2},
          eid = {staf1935},
        pages = {staf1935},
          doi = {10.1093/mnras/staf1935},
archivePrefix = {arXiv},
       eprint = {2506.15781},
 primaryClass = {astro-ph.GA},
       adsurl = {https://ui.adsabs.harvard.edu/abs/2026MNRAS.545f1935K}
}

@ARTICLE{Stone2020SSRv,
       author = {{Stone}, N.~C. and {Vasiliev}, E. and {Kesden}, M. and {Rossi}, E.~M. and {Perets}, H.~B. and {Amaro-Seoane}, P.},
        title = "{Rates of Stellar Tidal Disruption}",
      journal = {\ssr},
         year = 2020,
        month = mar,
       volume = {216},
       number = {3},
          eid = {35},
        pages = {35},
          doi = {10.1007/s11214-020-00651-4},
archivePrefix = {arXiv},
       eprint = {2003.08953},
 primaryClass = {astro-ph.HE},
       adsurl = {https://ui.adsabs.harvard.edu/abs/2020SSRv..216...35S}
}

@INCOLLECTION{Pau2022hgwa,
       author = {{Amaro Seoane}, Pau},
        title = "{The Gravitational Capture of Compact Objects by Massive Black Holes}",
    booktitle = {Handbook of Gravitational Wave Astronomy},
         year = 2022,
       editor = {{Bambi}, Cosimo and {Katsanevas}, Stavros and {Kokkotas}, Konstantinos D.},
          eid = {17},
        pages = {17},
          doi = {10.1007/978-981-15-4702-7_17-1},
       adsurl = {https://ui.adsabs.harvard.edu/abs/2022hgwa.bookE..17A}
}

@ARTICLE{Alexander2005PhR,
       author = {{Alexander}, Tal},
        title = "{Stellar processes near the massive black hole in the Galactic center [review article]}",
      journal = {\physrep},
         year = 2005,
        month = nov,
       volume = {419},
       number = {2-3},
        pages = {65-142},
          doi = {10.1016/j.physrep.2005.08.002},
archivePrefix = {arXiv},
       eprint = {astro-ph/0508106},
 primaryClass = {astro-ph},
       adsurl = {https://ui.adsabs.harvard.edu/abs/2005PhR...419...65A}
}

@ARTICLE{Merritt2013,
       author = {{Merritt}, David},
        title = "{Loss-cone dynamics}",
      journal = {Classical and Quantum Gravity},
         year = 2013,
        month = dec,
       volume = {30},
       number = {24},
          eid = {244005},
        pages = {244005},
          doi = {10.1088/0264-9381/30/24/244005},
archivePrefix = {arXiv},
       eprint = {1307.3268},
 primaryClass = {astro-ph.GA},
       adsurl = {https://ui.adsabs.harvard.edu/abs/2013CQGra..30x4005M}
}

@ARTICLE{Tomar2026PhRvD,
       author = {{Tomar}, Yashvardhan and {Hopkins}, Philip F. and {Kremer}, Kyle},
        title = "{Suppressed capture and merger rates in AGN}",
      journal = {\prd},
         year = 2026,
        month = mar,
       volume = {113},
       number = {6},
          eid = {063036},
        pages = {063036},
          doi = {10.1103/gcwt-rmml},
archivePrefix = {arXiv},
       eprint = {2601.02487},
 primaryClass = {astro-ph.HE},
       adsurl = {https://ui.adsabs.harvard.edu/abs/2026PhRvD.113f3036T}
}

@ARTICLE{Sirko2003,
       author = {{Sirko}, Edwin and {Goodman}, Jeremy},
        title = "{Spectral energy distributions of marginally self-gravitating quasi-stellar object discs}",
      journal = {\mnras},
         year = 2003,
        month = may,
       volume = {341},
       number = {2},
        pages = {501-508},
          doi = {10.1046/j.1365-8711.2003.06431.x},
archivePrefix = {arXiv},
       eprint = {astro-ph/0209469},
 primaryClass = {astro-ph},
       adsurl = {https://ui.adsabs.harvard.edu/abs/2003MNRAS.341..501S}
}

@ARTICLE{Whitehead2025MNRAS,
       author = {{Whitehead}, Henry and {Rowan}, Connar and {Kocsis}, Bence},
        title = "{Hydrodynamic simulations of black hole evolution in AGN discs II: inclination damping for partially embedded satellites}",
      journal = {\mnras},
         year = 2025,
        month = nov,
       volume = {543},
       number = {4},
        pages = {3768-3782},
          doi = {10.1093/mnras/staf1686},
archivePrefix = {arXiv},
       eprint = {2505.23899},
 primaryClass = {astro-ph.HE},
       adsurl = {https://ui.adsabs.harvard.edu/abs/2025MNRAS.543.3768W}
}

@ARTICLE{LYP2024ApJ,
       author = {{Li}, Ya-Ping and {Chen}, Yi-Xian and {Lin}, Douglas N.~C.},
        title = "{Concurrent Accretion and Migration of Giant Planets in Their Natal Disks with Consistent Accretion Torque}",
      journal = {\apj},
         year = 2024,
        month = aug,
       volume = {971},
       number = {2},
          eid = {130},
        pages = {130},
          doi = {10.3847/1538-4357/ad5a06},
archivePrefix = {arXiv},
       eprint = {2406.12716},
 primaryClass = {astro-ph.EP},
       adsurl = {https://ui.adsabs.harvard.edu/abs/2024ApJ...971..130L}
}

@ARTICLE{Tanaka2002ApJ,
       author = {{Tanaka}, Hidekazu and {Takeuchi}, Taku and {Ward}, William R.},
        title = "{Three-Dimensional Interaction between a Planet and an Isothermal Gaseous Disk. I. Corotation and Lindblad Torques and Planet Migration}",
      journal = {\apj},
         year = 2002,
        month = feb,
       volume = {565},
       number = {2},
        pages = {1257-1274},
          doi = {10.1086/324713},
       adsurl = {https://ui.adsabs.harvard.edu/abs/2002ApJ...565.1257T}
}

@ARTICLE{Tanaka2004ApJ,
       author = {{Tanaka}, Hidekazu and {Ward}, William R.},
        title = "{Three-dimensional Interaction between a Planet and an Isothermal Gaseous Disk. II. Eccentricity Waves and Bending Waves}",
      journal = {\apj},
         year = 2004,
        month = feb,
       volume = {602},
       number = {1},
        pages = {388-395},
          doi = {10.1086/380992},
       adsurl = {https://ui.adsabs.harvard.edu/abs/2004ApJ...602..388T}
}

@ARTICLE{Goldreich1979ApJ,
       author = {{Goldreich}, P. and {Tremaine}, S.},
        title = "{The excitation of density waves at the Lindblad and corotation resonances by an external potential.}",
      journal = {\apj},
         year = 1979,
        month = nov,
       volume = {233},
        pages = {857-871},
          doi = {10.1086/157448},
       adsurl = {https://ui.adsabs.harvard.edu/abs/1979ApJ...233..857G}
}

@ARTICLE{wmy2024ApJ,
       author = {{Wang}, Mengye and {Ma}, Yiqiu and {Wu}, Qingwen and {Jiang}, Ning},
        title = "{An Explanation for the Overrepresentation of Tidal Disruption Events in Post-starburst Galaxies}",
      journal = {\apj},
         year = 2024,
        month = jan,
       volume = {960},
       number = {1},
          eid = {69},
        pages = {69},
          doi = {10.3847/1538-4357/ad0bfb},
archivePrefix = {arXiv},
       eprint = {2311.07040},
 primaryClass = {astro-ph.HE},
       adsurl = {https://ui.adsabs.harvard.edu/abs/2024ApJ...960...69W}
}

@ARTICLE{wangyh2024ApJ,
       author = {{Wang}, Yihan and {Lin}, Douglas N.~C. and {Zhang}, Bing and {Zhu}, Zhaohuan},
        title = "{Changing-look Active Galactic Nuclei Behavior Induced by Disk-captured Tidal Disruption Events}",
      journal = {\apjl},
         year = 2024,
        month = feb,
       volume = {962},
       number = {1},
          eid = {L7},
        pages = {L7},
          doi = {10.3847/2041-8213/ad20e5},
archivePrefix = {arXiv},
       eprint = {2310.00038},
 primaryClass = {astro-ph.HE},
       adsurl = {https://ui.adsabs.harvard.edu/abs/2024ApJ...962L...7W}
}

@ARTICLE{Goldreich1978,
       author = {{Goldreich}, P. and {Tremaine}, S.},
        title = "{The excitation and evolution of density waves.}",
      journal = {\apj},
         year = 1978,
        month = jun,
       volume = {222},
        pages = {850-858},
          doi = {10.1086/156203},
       adsurl = {https://ui.adsabs.harvard.edu/abs/1978ApJ...222..850G}
}

@ARTICLE{Lei2026ApJ,
       author = {{Lei}, Xiangli and {Wu}, Qingwen and {Li}, Ya-Ping and {Lei}, Wei-Hua},
        title = "{Simulations of Tidal Disruption of Supernova in Galaxy Nuclear Region: A Novel Model for Ambiguous Nuclear Transients}",
      journal = {\apjl},
         year = 2026,
        month = jan,
       volume = {996},
       number = {1},
          eid = {L2},
        pages = {L2},
          doi = {10.3847/2041-8213/ae25f3},
archivePrefix = {arXiv},
       eprint = {2509.11186},
 primaryClass = {astro-ph.HE},
       adsurl = {https://ui.adsabs.harvard.edu/abs/2026ApJ...996L...2L}
}

@ARTICLE{wangyh2024arXiv,
       author = {{Wang}, Yihan and {Graham}, Matthew J. and {Ford}, K.~E. Saavik and {McKernan}, Barry and {Ryu}, Taeho and {Stern}, Daniel},
        title = "{Conditions for Changing-Look AGNs from Accretion Disk-Induced Tidal Disruption Events}",
      journal = {arXiv e-prints},
         year = 2024,
        month = jun,
          eid = {arXiv:2406.12096},
        pages = {arXiv:2406.12096},
          doi = {10.48550/arXiv.2406.12096},
archivePrefix = {arXiv},
       eprint = {2406.12096},
 primaryClass = {astro-ph.HE},
       adsurl = {https://ui.adsabs.harvard.edu/abs/2024arXiv240612096W}
}

@ARTICLE{Tagawa2020ApJ,
       author = {{Tagawa}, Hiromichi and {Haiman}, Zolt{\'a}n and {Kocsis}, Bence},
        title = "{Formation and Evolution of Compact-object Binaries in AGN Disks}",
      journal = {\apj},
         year = 2020,
        month = jul,
       volume = {898},
       number = {1},
          eid = {25},
        pages = {25},
          doi = {10.3847/1538-4357/ab9b8c},
archivePrefix = {arXiv},
       eprint = {1912.08218},
 primaryClass = {astro-ph.GA},
       adsurl = {https://ui.adsabs.harvard.edu/abs/2020ApJ...898...25T}
}

@ARTICLE{Macleod2020ApJ,
       author = {{MacLeod}, Morgan and {Lin}, Douglas N.~C.},
        title = "{The Effect of Star-Disk Interactions on Highly Eccentric Stellar Orbits in Active Galactic Nuclei: A Disk Loss Cone and Implications for Stellar Tidal Disruption Events}",
      journal = {\apj},
         year = 2020,
        month = feb,
       volume = {889},
       number = {2},
          eid = {94},
        pages = {94},
          doi = {10.3847/1538-4357/ab64db},
archivePrefix = {arXiv},
       eprint = {1909.09645},
 primaryClass = {astro-ph.SR},
       adsurl = {https://ui.adsabs.harvard.edu/abs/2020ApJ...889...94M}
}

@ARTICLE{McKernan2012,
       author = {{McKernan}, B. and {Ford}, K.~E.~S. and {Lyra}, W. and {Perets}, H.~B.},
        title = "{Intermediate mass black holes in AGN discs - I. Production and growth}",
      journal = {\mnras},
         year = 2012,
        month = sep,
       volume = {425},
       number = {1},
        pages = {460-469},
          doi = {10.1111/j.1365-2966.2012.21486.x},
archivePrefix = {arXiv},
       eprint = {1206.2309},
 primaryClass = {astro-ph.GA},
       adsurl = {https://ui.adsabs.harvard.edu/abs/2012MNRAS.425..460M}
}

@ARTICLE{Rowan2025MNRAS,
       author = {{Rowan}, Connar and {Whitehead}, Henry and {Fabj}, Gaia and {Kirkeberg}, Philip and {Pessah}, Martin E. and {Kocsis}, Bence},
        title = "{Hydrodynamic simulations of black hole evolution in AGN discs ─ I. Orbital alignment of highly inclined satellites}",
      journal = {\mnras},
         year = 2025,
        month = oct,
       volume = {543},
       number = {1},
        pages = {132-145},
          doi = {10.1093/mnras/staf1449},
archivePrefix = {arXiv},
       eprint = {2505.23739},
 primaryClass = {astro-ph.HE},
       adsurl = {https://ui.adsabs.harvard.edu/abs/2025MNRAS.543..132R}
}

@ARTICLE{Kupper2011MNRAS,
       author = {{K{\"u}pper}, Andreas H.~W. and {Maschberger}, Thomas and {Kroupa}, Pavel and {Baumgardt}, Holger},
        title = "{Mass segregation and fractal substructure in young massive clusters - I. The McLuster code and method calibration}",
      journal = {\mnras},
         year = 2011,
        month = nov,
       volume = {417},
       number = {3},
        pages = {2300-2317},
          doi = {10.1111/j.1365-2966.2011.19412.x},
archivePrefix = {arXiv},
       eprint = {1107.2395},
 primaryClass = {astro-ph.GA},
       adsurl = {https://ui.adsabs.harvard.edu/abs/2011MNRAS.417.2300K}
}

@ARTICLE{Kroupa2001MNRAS,
       author = {{Kroupa}, Pavel},
        title = "{On the variation of the initial mass function}",
      journal = {\mnras},
         year = 2001,
        month = apr,
       volume = {322},
       number = {2},
        pages = {231-246},
          doi = {10.1046/j.1365-8711.2001.04022.x},
archivePrefix = {arXiv},
       eprint = {astro-ph/0009005},
 primaryClass = {astro-ph},
       adsurl = {https://ui.adsabs.harvard.edu/abs/2001MNRAS.322..231K}
}

@ARTICLE{Lee2025ApJ,
       author = {{Lee}, Seungjae and {Lee}, Hyung Mok and {Kim}, Ji-hoon and {Spurzem}, Rainer and {Hong}, Jongsuk and {Chung}, Eunwoo},
        title = "{Formation and Evolution of Compact Binaries Containing Intermediate-mass Black Holes in Dense Star Clusters}",
      journal = {\apj},
         year = 2025,
        month = jul,
       volume = {988},
       number = {1},
          eid = {15},
        pages = {15},
          doi = {10.3847/1538-4357/adde52},
archivePrefix = {arXiv},
       eprint = {2503.22109},
 primaryClass = {astro-ph.GA},
       adsurl = {https://ui.adsabs.harvard.edu/abs/2025ApJ...988...15L}
}

@ARTICLE{WangLong2020MNRAS493,
       author = {{Wang}, Long and {Nitadori}, Keigo and {Makino}, Junichiro},
        title = "{A slow-down time-transformed symplectic integrator for solving the few-body problem}",
      journal = {\mnras},
         year = 2020,
        month = apr,
       volume = {493},
       number = {3},
        pages = {3398-3411},
          doi = {10.1093/mnras/staa480},
archivePrefix = {arXiv},
       eprint = {2002.07938},
 primaryClass = {astro-ph.EP},
       adsurl = {https://ui.adsabs.harvard.edu/abs/2020MNRAS.493.3398W}
}

@ARTICLE{WangLong2020MNRAS,
       author = {{Wang}, Long and {Iwasawa}, Masaki and {Nitadori}, Keigo and {Makino}, Junichiro},
        title = "{PETAR: a high-performance N-body code for modelling massive collisional stellar systems}",
      journal = {\mnras},
         year = 2020,
        month = sep,
       volume = {497},
       number = {1},
        pages = {536-555},
          doi = {10.1093/mnras/staa1915},
archivePrefix = {arXiv},
       eprint = {2006.16560},
 primaryClass = {astro-ph.IM},
       adsurl = {https://ui.adsabs.harvard.edu/abs/2020MNRAS.497..536W}
}

@ARTICLE{Just2012ApJ,
       author = {{Just}, Andreas and {Yurin}, Denis and {Makukov}, Maxim and {Berczik}, Peter and {Omarov}, Chingis and {Spurzem}, Rainer and {Vilkoviskij}, Emmanuil Y.},
        title = "{Enhanced Accretion Rates of Stars on Supermassive Black Holes by Star-Disk Interactions in Galactic Nuclei}",
      journal = {\apj},
         year = 2012,
        month = oct,
       volume = {758},
       number = {1},
          eid = {51},
        pages = {51},
          doi = {10.1088/0004-637X/758/1/51},
archivePrefix = {arXiv},
       eprint = {1208.4954},
 primaryClass = {astro-ph.CO},
       adsurl = {https://ui.adsabs.harvard.edu/abs/2012ApJ...758...51J}
}

@ARTICLE{Kennedy2016MNRAS,
       author = {{Kennedy}, Gareth F. and {Meiron}, Yohai and {Shukirgaliyev}, Bekdaulet and {Panamarev}, Taras and {Berczik}, Peter and {Just}, Andreas and {Spurzem}, Rainer},
        title = "{Star-disc interaction in galactic nuclei: orbits and rates of accreted stars}",
      journal = {\mnras},
         year = 2016,
        month = jul,
       volume = {460},
       number = {1},
        pages = {240-255},
          doi = {10.1093/mnras/stw908},
archivePrefix = {arXiv},
       eprint = {1604.05309},
 primaryClass = {astro-ph.GA},
       adsurl = {https://ui.adsabs.harvard.edu/abs/2016MNRAS.460..240K}
}

@ARTICLE{Iwasawa2020PASJ,
       author = {{Iwasawa}, Masaki and {Namekata}, Daisuke and {Nitadori}, Keigo and {Nomura}, Kentaro and {Wang}, Long and {Tsubouchi}, Miyuki and {Makino}, Junichiro},
        title = "{Accelerated FDPS: Algorithms to use accelerators with FDPS}",
      journal = {\pasj},
         year = 2020,
        month = feb,
       volume = {72},
       number = {1},
          eid = {13},
        pages = {13},
          doi = {10.1093/pasj/psz133},
archivePrefix = {arXiv},
       eprint = {1907.02290},
 primaryClass = {astro-ph.IM},
       adsurl = {https://ui.adsabs.harvard.edu/abs/2020PASJ...72...13I}
}

@ARTICLE{Iwasawa2016PASJ,
       author = {{Iwasawa}, Masaki and {Tanikawa}, Ataru and {Hosono}, Natsuki and {Nitadori}, Keigo and {Muranushi}, Takayuki and {Makino}, Junichiro},
        title = "{Implementation and performance of FDPS: a framework for developing parallel particle simulation codes}",
      journal = {\pasj},
         year = 2016,
        month = aug,
       volume = {68},
       number = {4},
          eid = {54},
        pages = {54},
          doi = {10.1093/pasj/psw053},
archivePrefix = {arXiv},
       eprint = {1601.03138},
 primaryClass = {astro-ph.IM},
       adsurl = {https://ui.adsabs.harvard.edu/abs/2016PASJ...68...54I}
}
\bibliographystyle{aasjournal}

\end{document}